\documentclass[a4paper,fleqn,usenatbib]{mnras}

\usepackage{newtxtext,newtxmath}

\usepackage[T1]{fontenc}
\usepackage{ae,aecompl}

\usepackage{epsfig}
\usepackage{amsmath}
\usepackage{amssymb}
\usepackage{natbib}
\usepackage{multirow}
\usepackage{graphicx}\graphicspath{{graphics/}}
\usepackage{color}
\usepackage[normalem]{ulem}
\usepackage{epsfig}
\usepackage{mathtools}
\usepackage{gensymb}
\usepackage{caption}
\usepackage{arydshln}
\usepackage{float}
\usepackage{hyperref}
\usepackage{nccmath}
\usepackage{subfigure}
\usepackage{tasks}
\settasks{
	counter-format=(tsk[r]),
	label-width=4ex}

\newcommand{\RM}[1]{\MakeUppercase{\romannumeral #1{}}}
\newcommand{\RMM}[1]{\MakeLowercase{\romannumeral #1{}}}

\title[Kinematics of Simulated Galaxies \RM{2}]{Kinematics of Simulated Galaxies \RM{2}: Probing the Stellar Kinematics of Galaxies out to Large Radii}

\author[Schulze et al.]{Felix Schulze$^{1,2}$\thanks{E-mail: fschulze@usm.lmu.de}, Rhea-Silvia Remus$^{1,3}$, Klaus Dolag$^{1,4}$, Sabine Bellstedt$^{5,6}$, \newauthor Andreas Burkert$^{1,2}$, and Duncan A. Forbes$^{5}$
\\
$^{1}$ Universit\"ats-Sternwarte M\"unchen, Scheinerstr.\ 1, D-81679 M\"unchen, Germany\\
$^{2}$ Max Planck Institute for Extraterrestrial Physics, Giessenbachstra{\ss}e 1, D-85748 Garching, Germany\\
$^{3}$ Canadian Institute for Theoretical Astrophysics, 60 St. George Street, University of Toronto, Toronto ON M5S 3H8, Canada\\
$^{4}$ Max Planck Institut for Astrophysics, D-85748 Garching, Germany\\
$^{5}$ Centre for Astrophysics and Supercomputing, Swinburne University of Technology, Hawthorn VIC 3122, Australia\\
$^{6}$ ICRAR, The University of Western Australia, 7 Fairway, Crawley WA 6009, Australia
}

\date{Accepted XXX. Received YYY; in original form ZZZ}

\pubyear{2019}

\begin{document}
\label{firstpage}
\pagerange{\pageref{firstpage}--\pageref{lastpage}}
\maketitle

\begin{abstract}
We investigate the stellar kinematics of a sample of galaxies extracted from the hydrodynamic cosmological \textit{Magneticum Pathfinder} simulations out to $5$ half-mass radii. We construct differential radial stellar spin profiles quantified by the observationally widely used $\lambda_\mathrm{R}$ and the closely related $(V/\sigma)$ parameters. We find three characteristic profile shapes: profiles exhibiting a (\RMM{1}) peak within $2.5$ half-mass radii and a subsequent decrease (\RMM{2}) continuous increase that plateaus at larger radii typically with a high amplitude (\RMM{3}) completely flat behaviour typically with low amplitude, in agreement with observations. This shows that the kinematic state of the stellar component can vary significantly with radius, suggesting a distinct interplay between in-situ star formation and ex-situ accretion of stars. Following the evolution of our sample through time, we provide evidence that the accretion history of galaxies with decreasing profiles is dominated by the anisotropic accretion of low mass satellites that get disrupted beyond $ \sim 2.0$ half-mass radii, building up a stellar halo with non-ordered motion while maintaining the central rotation already present at $z=2$. In fact, at $z=2$ decreasing profiles are the predominant profile class. Hence, we can predict a distinct formation pathway for galaxies with a decreasing profile and show that the centre resembles an old embedded disk. Furthermore, we show that the radius of the kinematic transition provides a good estimation for the transition radius from in-situ stars in the centre to accreted stars in the halo.
\end{abstract}





\section{Introduction}

Within the $\Lambda \mathrm{CDM}$ paradigm, dark matter structures merge hierarchically at high redshift, leading to the build-up of deep potential wells that efficiently funnel gas into the centre of halos, proceeding the galaxy formation via primordial insitu star formation at the very centre of a galaxy. Subsequent to this dissipative epoch, below $z \approx 2$ the dominant process shaping the properties of galaxies is galactic mergers \citep{2010ApJ...725.2312O,2013ApJ...768L..28H}. A main driver of the relative importance of dissipative and merging processes is the stellar mass \citep{2018MNRAS.478.3994C}. While this multi phase scenario represents a suitable picture of galaxy formation on cosmological time-scales, the mechanisms determining the detailed inner baryonic structure of galaxies are still not completely understood \citep{2017ARA&A..55...59N}.
\par
The broad morphological distinction into late-type galaxies (LTGs) and early-type galaxies (ETGs) is mainly driven by the low-redshift evolution of galaxies. While late-type galaxies are expected to experience a quiet formation pathway, mostly driven by internal secular processes that leave the disc structure in the centre intact, early-type galaxies are subject to a complex interplay between environmental processes like mergers, tidal striping, ram-pressure stripping, harassment and strangulation.
\par
Early observations based on photometry perceived ETGs as fairly simple objects without significant internal structure. In contrast to the accepted picture at that time, \citet{1988A&AS...74..385B} showed that isophotes of ETGs differ from ellipses and can be either 'boxy' or 'discy'. Together with the discovery that discy ETGs seemed to rotate more rapidly than boxy ETGs, the correlation between the isophotal shape and the central surface density slope led to a revision of the formation picture of ETGs \citep{1996ApJ...464L.119K,1997AJ....114.1771F}.
\par
Due to the availability of integral-field observations, this picture has been significantly advanced during the past decade, providing highly-resolved two-dimensional maps of velocities and velocity dispersions (as well as other properties). In pioneering works the SAURON \citep{2001MNRAS.326...23B} and $\mathrm{ATLAS^{3D}}$ \citep{2011MNRAS.413..813C} surveys revealed a dichotomy in the central stellar kinematics of ETGs: Fast rotating ETGs show a globally regular rotation velocity pattern in agreement with an inclined rotating disc, while slow rotating ETGs are dispersion dominated with no sign of global rotational support. Later-on, these results have been strengthened and further extended by IFS surveys like CALIFA \citep{2012A&A...538A...8S}, SAMI \citep{2012MNRAS.421..872C}, SLUGGS \citep{2014ApJ...796...52B}, MaNGA \citep{2015AJ....149...77D}, and MASSIVE \citep{2014ApJ...795..158M}, complemented by results from numerical simulations exploring the origin of this kinematic dichotomy \citep[e.g.][]{2009MNRAS.397.1202J,2011MNRAS.416.1654B,2014MNRAS.444.3357N,2017MNRAS.468.3883P,2017ApJ...837...68C,2018MNRAS.480.4636S,2018ApJ...856..114C,2019MNRAS.484..869V}. Studies of the very central regions of slow rotators show a variety of kinematic subcomponents such as different types of kinematically distinct cores \citep{2006MNRAS.373..906M,2010ApJ...723..818H,2011MNRAS.414.2923K,2011MNRAS.414..888E,2013MNRAS.432.1768K,2017Galax...5...41S}. Those results are typically obtained by observations restricted to an aperture of $\sim 1$ effective radius ($R_{\mathrm{e}}$) due to the challenging task of observing the faint stellar halo of galaxies.
\par
Environmental effects play a fundamental role in determining the present-day properties of galaxies, especially for ETGs. Due to the short mixing time-scales in the centre of galaxies it is difficult to decipher the imprints of these effects in the central kinematics. Since the mixing time-scales in the stellar halos of galaxies are substantially longer than in the centre, stellar halos represent an excellent laboratory to probe the accretion histories of galaxies. Furthermore, the connection between the central and halo stellar kinematics encodes information about the formation pathway of galaxies.
\par
As a main driver of galaxy formation, galaxy mergers are one of the most important environmental effects. However, the impact of mergers at lower redshift is not trivial to predict. The morphology of (dry) major merger \footnote{Note, however, that wet major merger can also lead to the formation of a disk galaxy, see \citet{2017MNRAS.470.3946S}.} remnants naturally resembles a puffed-up spheroidal due to the predominant effect of violent relaxation, capable of drastically altering the global orbital configuration of the progenitors \citep{2010ApJ...723..818H,2014MNRAS.444.1475M}. In contrast, the mass accreted through dry minor mergers is in general deposited in the outskirts of the galaxy, leading to a less concentrated remnant, influencing the galaxy properties at larger radii beyond $\sim 1R_{\mathrm{e}}$ \citep{2012MNRAS.425.3119H,2012MNRAS.425..641L,2019MNRAS.487..318K}. Thus, galaxies are expected to display structural variations in radial profiles of properties such as the surface brightness \citep[Remus\&Forbes 2019, in prep]{2014MNRAS.443.1433D,2017A&A...603A..38S,2013MNRAS.434.3348C,2006MNRAS.365..747A}, age and metallicity \citep{2016ApJ...820..131E,2014MNRAS.442.1003P,2015MNRAS.451.2625P,2018MNRAS.479.4760F}, and stellar kinematics.
\par
Noticeable progress in understanding stellar halo kinematics has been made using spatially sparse tracer measurements beyond $1R_{\mathrm{e}}$ using planetary nebulae \citep{2009MNRAS.394.1249C,2009ApJ...691..228M,2018A&A...618A..94P} and globular clusters \citep{2001ApJ...563..135M,2011ApJS..197...33S,2011ApJ...736L..26A,2013MNRAS.428..389P}. Especially \citet{2009MNRAS.394.1249C} present pioneering kinematic measurements based on planetary nebulae out to $\sim 10R_{\mathrm{e}}$. While these studies represent an important advance in understanding the kinematics of the stellar halo, they are limited by a small sample size (some focusing on single galaxies), and therefore do not provide statistically representative results.
\par
Within the SAGES Legacy Unifying Globulars and GalaxieS (SLUGGS, \citet{2014ApJ...796...52B}) survey, \citet{2014ApJ...791...80A} compiled differential specific angular momentum profiles out to $\sim 2-4 R_{\mathrm{e}}$ for $22$ ETGs using multiple slitmask spectra (SKiMS). For a significant fraction of galaxies they found that the kinematic state observed within $1R_{\mathrm{e}}$ could change drastically at larger radii. While centrally slow rotating galaxies typically remain dispersion dominated in the outskirts, centrally fast rotating galaxies show profiles ranging from rapidly increasing to declining at larger radii. The declining profiles are interpreted as a sign for a kinematic subcomponent, i.e. an embedded disk. This result was later confirmed by \citet{2016MNRAS.457..147F} using a similar but slightly extended sample of $25$ ETGs. In contrast, drops in stellar spin reported by \citet{2014ApJ...786...23R} and \citet{2017MNRAS.471.4005B} were not as strong as reported for the SLUGGS galaxies. The discrepancies are mainly attributed to small sample sizes and selection biases. The physical meaning of embedded disks has been emphasised by \citet{2016MNRAS.457..320S} and \citet{2016ApJ...831..132G} who showed that they are a key component to correctly model galaxies with respect to photometry and the related black hole mass.
\par
In a first step to understand radially-varying kinematics, more recent studies investigated the connection between the modulations of the stellar spin profile and the visual morphology of galaxies. Within the CALIFA project, \citet{2016ASPC..505..133F} found a continuous sequence spanning high to low rotational support when going from later to earlier galaxy types at all radii out to $3 R_{\mathrm{e}}$. It is, however, not straight-forward to compare the results of this study to results from the SLUGGS survey since the authors calculate stellar spin profiles in an integrated manner rather than differential. Focusing only on ETGs, \citet{2017MNRAS.467.4540B} observed a clear difference in the local stellar spin profiles between lenticular (S0) and elliptical (E) galaxies within their sample of $28$ SLUGGS ETGs. While the vast majority of S0 galaxies show a rapidly rising profile with a plateau at larger radii, a significant fraction of ellipticals exhibit a completely flat or declining profile. A connection between the apparent morphology and the rotational support was also found by \citet{2018MNRAS.480.3105F} for a significantly larger galaxy sample including late-type galaxies. In one of the few theoretical studies, \citet{2014MNRAS.438.2701W} investigated the stellar kinematics of $42$ zoom simulations of galaxies out to $\sim 6R_{\mathrm{e}}$. Consistent with the observations, the simulation produces inclining, declining and constant radial angular momentum profiles.
\par
Very recently, the boundaries of observing stellar halo kinematics has been further pushed by \citet{2018A&A...618A..94P}. Using planetary nebulae as tracers for the underlying stellar velocity field, they observed stellar kinematics out to typically $\sim 6R_{\mathrm{e}}$, with a range of $[3R_{\mathrm{e}}-13R_{\mathrm{e}}]$. Interestingly, they found that the majority of centrally fast-rotating ETGs exhibit declining $V/\sigma$ profiles as well as declining velocity amplitudes. In contrast, slow rotators typically show higher rotational support in the stellar halo than within $1R_{\mathrm{e}}$. The authors conclude that the varying rotational support of fast rotators is due to a more or less prominent disk in the centre which fades towards larger radii. As these transitions occur usually at larger radii than most studies have been covering so far, a statistical significance in the appearance of the different radial behaviour could not yet be established and is one of the major goals of the work presented here.
\par
Therefore, this study uses data from the fully cosmological hydrodynamic \textit{Magneticum Pathfinder} simulations to investigate the stellar kinematics of massive galaxies out to $\sim 5$ half-mass radii ($R_{\mathrm{1/2}}$) in a statistical manner and understand the origin of the different radial behaviours present in observations with special emphasis on the declining stellar spin profiles. The paper is structured as follows: Details of the simulation and the methodology of the galaxy analysis as well as the sample selection are presented in Sec.~\ref{sec:Simulation_and_Analysis}. In Sec.~\ref{sec:radial_profiles_put_to_5Re} we conduct a first qualitative investigation of the radial stellar spin profiles and compare to recent observations. Furthermore, we introduce a new classification based on the shape of the spin profiles that gets explored in detail throughout the study. Therefore, Sec.~\ref{sec:Profile_Shape_Correlation_with_Galaxy_Properties} investigates the connection between the stellar large scale kinematics and fundamental galaxy properties like central stellar kinematics, stellar mass, and morphology. Exploiting the full power of the simulation, in Sec.~\ref{sec:formation_pathway} we follow the evolution of our sample through cosmic time and investigate the imprint of the accretion history on the stellar large scale kinematics. We conclude in Sec.~\ref{sec:Summary_and_Conclusion} with a summary and conclusion.


\section{Simulation and Analysis} \label{sec:Simulation_and_Analysis}

\subsection{The Magneticum Pathfinder Simulations}

For our analysis we study galaxies extracted from the cosmological hydrodynamical \textit{Magneticum\footnote{www.magneticum.org} Pathfinder} simulations, which are a collection of simulations of various box-sizes and resolutions. Sizes range from $18 \mathrm{Mpc/h}$ to $2688 \mathrm{Mpc/h}$ box side length, while resolutions cover a particle mass range of $10^{10} > m_{\mathrm{dm}} > 10^7 M_{\odot}/h$ for the dark matter and $10^9 > m_{\mathrm{gas}} > 10^6 M_{\odot}/h$ for gas particles. The simulations were performed with the Tree/SPH code GADGET-3 which is an extended version of GADGET-2 \citep{2005MNRAS.364.1105S, 2001NewA....6...79S} implementing updates in the SPH formulation regarding the treatment of viscosity and the used kernels \citep{2005MNRAS.364..753D, 2016MNRAS.455.2110B}.
\par
The simulation implements state-of-the-art models for a variety of baryonic physics needed for the self-consistent modelling of galaxy formation such as gas cooling and star formation \citep{2003MNRAS.339..289S}, black hole seeding, evolution and AGN feedback \citep{2005MNRAS.361..776S, 2014MNRAS.442.2304H,2015MNRAS.448.1504S} as well as stellar evolution and metal enrichment \citep{2007MNRAS.382.1050T}. Furthermore, it follows the thermal conduction similar to \citet{2004ApJ...606L..97D} but with $1/20\mathrm{th}$ of the classical Spitzer value \citep{1962pfig.book.....S} motivated by full MHD simulations including an anisotropic treatment of thermal conduction \citep{2014arXiv1412.6533A}. For more details on the baryonic physical models we refer to \citet{2015ApJ...812...29T}.
\par
The \textit{Magneticum Pathfinder} simulations have shown to successfully reproduce, and help to interpret, various observational results, such as pressure profiles of the intra-cluster medium \citep{2013A&A...550A.131P,2014ApJ...794...67M}, the predicted Zeldovich signal \citep{2016MNRAS.463.1797D}, galaxy cluster properties \citep{2017Galax...5...49R,2019MNRAS.488.5370L}, the properties of the AGN population \citep{2014MNRAS.442.2304H,2015MNRAS.448.1504S,2016MNRAS.458.1013S,2018MNRAS.481..341S}, the kinematic properties of galaxies \citep{2015ApJ...812...29T,2018MNRAS.480.4636S} and the dynamical properties of early-type galaxies \citep{2013ApJ...766...71R,2017MNRAS.472.4769T}.
\par
Throughout all simulated volumes and resolutions the simulations employ a standard $\Lambda$CDM cosmology with parameters adopted from the seven-year results of the Wilkinson Microwave Anisotropy Probe (WMAP7) \citep{2011ApJS..192...18K}, with $\Omega_{\mathrm{b}}=0.0451$, $\Omega_{\mathrm{M}}=0.272$ and $\Omega_{\Lambda}=0.728$ for baryons, matter and dark energy, respectively. Furthermore, the Hubble parameter is $h=0.704$ and the normalization of the fluctuation amplitude at $8 \mathrm{Mpc}$ is given by $\sigma_8=0.809$.
\par
The galaxies used in this work are extracted from Box4, a box with a side length of $48\mathrm{Mpc/h}$ at the currently highest resolution level, and initially contains $2 \times 576^3$ particles (gas and dark matter). The gravitational softening length is $\epsilon_{\mathrm{gas}}=\epsilon_{\mathrm{dm}}=1.4\mathrm{kpc/h}$ for dark matter and gas particles and $\epsilon_{\mathrm{*}}=0.7\mathrm{kpc/h}$ for stellar particles. The (initial) particle mass of dark matter and gas is $m_{\mathrm{DM}}=3.6 \times 10^7 M_{\odot}/h$ and $m_{\mathrm{gas}}=7.3 \times 10^6 M_{\odot}/h$, respectively. Since a gas particle can spawn up to $4$ stellar particles the mass of gas particles is not constant. Furthermore, the mass of stellar particles varies due to stellar wind losses, with an average mass of $M_*=2 \times 10^6 M_{\odot}/h$

\subsection{Sample Selection} \label{sec:sample_selection}

We select all halos identified by SUBFIND with a stellar mass $M_* > 2 \times 10^{10} M_{\odot}$ to ensure that the galaxy is sampled with a sufficient number of stellar particles. We find $1147$ halos in \textit{Magneticum} Box4 that obey this mass limit. In the next step we discard all galaxies with a stellar half-mass radius\footnote{The half-mass radius $R_{\mathrm{1/2}}$ is defined to be the radius of a three dimensional sphere containing half of the total stellar mass. Throughout this study $R_{\mathrm{1/2}}$ is considered to be equal to the observationally accessible effective radius $R_{\mathrm{e}}$.} $R_{\mathrm{1/2}} < 1.4~\mathrm{kpc~h^{-1}}$, corresponding to twice the stellar gravitational softening length for stars in the simulation. This selection criterion ensures an adequate spatial resolution, excluding galaxies dominated by the unresolved central region.
\par
This leaves us with a sample of $1132$ halos that have already been shown in previous work to successfully reproduce stellar kinematic properties within $1 R_{\mathrm{1/2}}$ (see \citealt{2018MNRAS.480.4636S} and \citealt{2019MNRAS.484..869V}).
 \par
Additionally, we discard galaxies from the sample whose stellar kinematics are not resolved beyond $3R_{\mathrm{1/2}}$ due to the lack of particles (see Sec. \ref{sec:construct_profile} for more details). After this step the sample contains $492$ objects, with $320$ of them beeing centrals and $172$ beeing satellites.
\par
In summary, we cover a mass range from $2 \times 10^{10} \mathrm{M_{\odot}}$ up to $\approx 1.7 \times 10^{12} \mathrm{M_{\odot}}$ with decreasing completeness towards higher and lower masses. At the high-mass end this is due to the limited box
-size that does not allow for the formation of a significant number of massive object with $M_* > 10^{12}$. Towards lower masses this is driven by the lack of particles in the galaxies leading to kinematic profiles that do not reach radii beyond $3R_{\mathrm{1/2}}$ and are therefore discarded from the sample due to our selection criteria.

\subsubsection{Kinematic Feature Classification}

\begin{figure}
		\centering
        \begin{center}
                \includegraphics[width=0.47\textwidth]{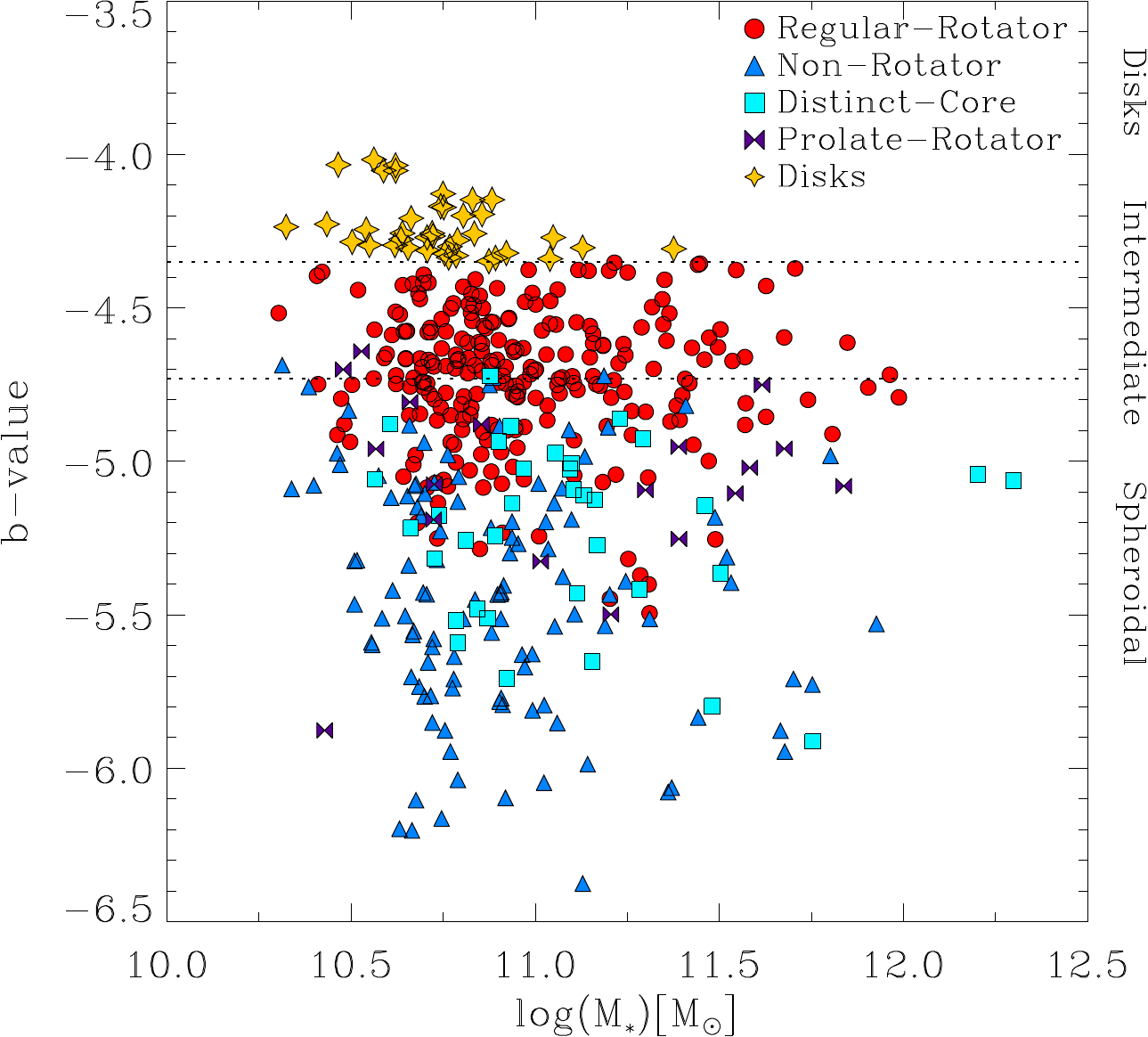}
        \end{center}
        \caption{The b-value as proxy for the morphology versus the stellar mass of the selected sample. The symbols discriminate Regular-Rotators (red circles), Non-Rotators (blue triangles), Distinct-Cores (cyan square), Prolate-Rotators (lilac bowtie), and Disks (yellow stars). The horizontal dashed lines indicate the threshold between Disks, Intermediates and Spheroidals as given in Sec.~\ref{sec:morph_class}.}
        {\label{fig:mass_b}}
\end{figure}
\par
In order to investigate the properties of our sample in more detail we apply a kinematic classification capturing specific kinematic features in the velocity maps of the central $1R_{\mathrm{1/2}}$ analogous to \citet{2018MNRAS.480.4636S} \citep[see also ][]{2011MNRAS.414.2923K}, and split our sample in five kinematic classes:
\begin{itemize}
	\item \textit{Regular-Rotator (RR):} The velocity map shows an ordered rotation pattern around the morphological minor axis.
	\item \textit{Non-Rotator (NR):} The velocity map does not show any signs of ordered motion and low level velocities.
	\item \textit{Distinct-Core (DC):} The velocity map shows a kinematically decoupled structure in the centre. This includes especially misaligned rotating structures with respect to the surrounding rotating galaxy, and rotating structures embedded in the centre of a non-rotating galaxy.
	\item \textit{Prolate-Rotator (PR):} The velocity map shows ordered rotation around the morphological major axis.
	\item \textit{Disks (D):} The velocity map shows a high degree of ordered rotation and a b-value (Eqn.~\ref{eqn:b}) larger than $-4.35$, which is the threshold for pure disk galaxies determined in \citet{2015ApJ...812...29T}.
\end{itemize}
\par
Discarding all galaxies that can not be classified into one of the above kinematic classes leaves us with a final sample of $450$ objects. Not classifiable galaxies typically exhibit unstructured velocity maps often caused by environmental effects, as tidal interactions with other objects. This selection process ensures that our sample is as clean as possible with regard to peculiar objects and interacting galaxies.

\subsubsection{Morphological Classification Using the b-value} \label{sec:morph_class}

To classify our sample by morphology we resort to the fundamental findings by \citet{1983IAUS..100..391F}: investigating the stellar mass $\mathrm{log(M_*)}$ and the specific angular momentum $\mathrm{log(j_*)}$ of galaxies they found that LTGs and ETGs follow a parallel sequence. Studies based on larger samples confirmed a continuous parallel sequence from pure disks to pure bulges in this plane corresponding to a shift of the bulge-to-total ratio from $0$ to $1$ \citep{2013ApJ...769L..26F,2014ApJ...784...26O,2016MNRAS.463..170C,2018ApJ...868..133F,2018ApJ...860...37S}. Furthermore, these results are in agreement with predictions from cosmological simulations \citet{2015ApJ...812...29T,2015ApJ...804L..40G,2016MNRAS.460.4466Z,2017MNRAS.464.3850L}. \citet{2015ApJ...812...29T} hence introduced the so called b-value as a proxy for morphology:
\begin{ceqn}
\begin{equation} \label{eqn:b}
	b=\mathrm{log_{10}}\left(\frac{j_*}{\mathrm{kpc~km/s}}\right)-\frac{2}{3}\mathrm{log_{10}}\left(\frac{M_*}{M_{\odot}}\right)
\end{equation}
\end{ceqn}
As found by \citet{2012ApJS..203...17R} and \citet{2015ApJ...812...29T}, galaxies with $b \approx -4$ are disc-like galaxies, followed by a smooth transition to lenticular and elliptical galaxies with decreasing b-value.
\par
Following \citet{2015ApJ...812...29T} we classify our galaxies according to their b-value as:
\begin{itemize}
	\item Disk: $b > -4.35$
	\item Intermediate: $-4.73 < b < -4.35$
	\item Spheroidal: $b < -4.73$
\end{itemize}
\par
Fig. \ref{fig:mass_b} summarizes the properties of our sample with regard to the introduced classifications. It shows the stellar mass of each galaxy versus the b-value. The symbols and colours distinguish RRs (red circles), NRs (blue triangles), DCs (cyan square), PRs (lilac bowtie), and Ds (yellow stars). The horizontal dashed lines separate Disks, Intermediates and Spheroids.
\par
The final galaxy sample studied in this work contains $39~(9\%)$ Disks, $140~(31\%)$) Intermediates and $271~(60\%)$ Spheriodals. Furthermore, the sample comprises $245~(54\%)$ Regular-Rotators, $113~(25\%)$ Non-Rotators, $35~(8\%)$ Distinct-Cores, $18~(4\%)$ Prolate-Rotators, and $39~(9\%)$ Disks. In comparison to the statistically larger sample investigated in \citet{2018MNRAS.480.4636S} the distribution is quite similar, demonstrating the completeness with regard to the kinematic classes. As expected the Regular-Rotators populate the high b-values range, while the Non-Rotators and the complex kinematic structures exhibit lower b-values in line with \citet{2018MNRAS.480.4636S}.

\subsection{Stellar Spin Analysis} \label{sec:kinematic_parameters}

To investigate the stellar kinematics of galaxies in the \textit{Magneticum} simulations, we construct two-dimensional spatially resolved line-of-sight velocity and velocity dispersion maps, and analyse their properties. This enables us to directly compare and apply the results to the wealth of observational IFS studies that have been carried out during the past decade and will be conducted in the future. Therefore, the construction of the kinematic maps should be as similar as possible to the methods used for the observations, while taking the properties and limitation of simulated data into account. The methods used in this study resemble the methods outlined in \citet{2018MNRAS.480.4636S}.
\par
Due to the limited mass resolution of the simulation, low particle numbers can cause statistical errors when sampling stellar particles onto a grid. To avoid statistical noise, we apply the following procedure to each galaxy in our sample: For a given spatial region and projection of interest the stellar particles are sampled onto a simple rectangular grid with a resolution comparable to modern IFS instruments of $0.3 \mathrm{kpc}$. Subsequently, we apply a Centroidal Voronoi Tessellation (CVT), which ensures a sufficient particle number per cell while maintaining an adequate spatial resolution. To construct the final kinematic maps the mean velocity
\begin{ceqn}
\begin{equation}
    \overline V_i=\frac{\sum_{j=1}^{N_c} V_j}{N_c}
\end{equation}
and the velocity dispersion
\begin{equation}
    \sigma_i=\sqrt{\frac{\sum_{j=1}^{N_c} V_j^2}{N_c}-\left(\frac{\sum_{j=1}^{N_c} V_j}{N_c}\right)^2}
\end{equation}
\end{ceqn}
are calculated within each CVT cell. Here, $V_j$ is the particle velocity, and the sum runs over all $N_c$ particles within the cell.

\subsubsection{Calculating Central Cumulative $\lambda_{\mathrm{R_{1/2}}}$}

During the course of this paper we will use the $\lambda_{\mathrm{R}}$ parameter which was introduced by \citet{2007MNRAS.379..401E} and has been extensively studied in great detail since then for many sets of observations \citep{2011MNRAS.414..888E,2013MNRAS.432.1768K,2014MNRAS.443..485F,2016MNRAS.463..170C,2017MNRAS.472.1272V,2017ApJ...844...59B,2017ApJ...851L..33G,2017MNRAS.471.1428V,2019arXiv191006236F} and simulations \citep{2009MNRAS.397.1202J,2011MNRAS.416.1654B,2014MNRAS.444.3357N,2014MNRAS.444.1475M,2017MNRAS.468.3883P,2017ApJ...837...68C,2018ApJ...856..114C,2018MNRAS.480.4636S}. For a given set of observed kinematic maps it is defined as
\begin{equation}
    \lambda_{R}=\frac{\sum_{i=1}^{N_{p}}F_{i} ~ R_{i} ~ |\overline V_{i}|}{\sum_{i=1}^{N_{p}}F_{i} ~ R_{i} ~ \sqrt{\overline V_{i}^2+\sigma_{i}^2}},
    \label{eq:lambda_r_obs}
\end{equation}
with the sum running over all pixels in the considered field of view. $F_i$, $R_i$, $\left|\overline V_i \right|$, and $\sigma_i$ are the flux, projected distance to the galaxy centre, mean stellar velocity, and velocity dispersion of the $\mathrm{i^{th}}$ photometric bin, respectively. For simulated data we replace the fluxes in Eq.~\ref{eq:lambda_r_obs} by stellar masses, assuming a constant mass-to-light ratio, as been done in various former theoretical studies \citep{2011MNRAS.416.1654B,2014MNRAS.444.3357N,2014MNRAS.444.1475M,2017MNRAS.468.3883P,2018MNRAS.480.4636S}.
\par
$\lambda_{\mathrm{R}}$ obviously depends on the projection direction of the velocity map. In practice, we choose the morphological edge-on projection, since this maximises $\lambda_{\mathrm{R}}$. Furthermore, the spatial region over which the summation in Eq.~\ref{eq:lambda_r_obs} is done, is given by an ellipse with the corresponding axis ratio $q_{\mathrm{morph}}$ of the galaxy and an area of $A_{\mathrm{ellipse}}=\pi~R_{\mathrm{1/2}}$. This is indicated by appending the subscript '1/2' to the parameter name $\lambda_{\mathrm{R_{1/2}}}$.

\subsubsection{Constructing Local Radial $\lambda(R)$ Profiles} \label{sec:construct_profile}

Since we aim to quantify the local variations of the kinematic state of galaxies, we calculate radial differential $\lambda(R)$ profiles for our sample. Calculating the profile in a differential manner, instead of cumulative, helps to capture kinematic transitions and the related profile gradients in more detail \citep{2017MNRAS.467.4540B}. Furthermore, using $\lambda(R)$ as proxy for the local stellar spin ensures a proper comparability to results from current and future IFS observations. 
\par
In order to determine the local $\lambda(R)$, we construct kinematic IFS mock observations of the line-of-sight velocity and velocity dispersion out to $5R_{1/2}$, using a CVT as described above. The velocities are transformed into the centre-of-mass system of the stars inside $1R_{1/2}$. Throughout this study we will refer to the local $\lambda$ by $\lambda(R)$.
\par
The CVT maps are the maximal statistically reliable information that can be extracted from the particle data. We intentionally refuse to apply interpolation techniques to the map that are not physically motivated since they might introduce artificial values. Therefore, we proceed by binning the CVT maps onto a rectangular grid. This artificially increases the resolution of the maps without introducing artificial values.
Then we calculate $\lambda(R)$ within elliptical annuli:
\begin{ceqn}
\begin{equation}
	\lambda(R)=\frac{\sum_{i=1}^{N_{p}} R_{i,c} ~ |\overline V_{i}|}{\sum_{i=1}^{N_{p}} R_{i,c} ~ \sqrt{\overline V_{i}^2+\sigma_{i}^2}},
	\label{eq:lambda_rad}
\end{equation}
\end{ceqn}
where summations run over all pixels within the annulus. Here, $R_{i}$, $V_{i}$ and $\sigma_{i}$ are the circularised projected distance to the galaxy centre, the mean stellar line-of-sight velocity and the velocity dispersion of the $\mathrm{i^{th}}$ photometric bin, respectively. The circularised distance is given by $R_{\mathrm{c}}$:
\begin{ceqn}
\begin{equation}
R_{\mathrm{c}}=\sqrt{x^2~q_{\mathrm{morph}}~+~y^2/q_{\mathrm{morph}}},
\end{equation}
\end{ceqn}
where $q_{\mathrm{morph}}$ is the morphological axis ratio of the galaxy. This transformation maps a given ellipse onto a circle with radius $R_{\mathrm{c}}$. The borders of the annuli are determined by demanding a minimum of $1500$ particles per radial bin. This ensures a proper equal statistical weighting of each annuli from the particle data. To reduce small scale noise the resulting radial profile is smoothed using an adaptive Gaussian kernel which uses $6$ neighbouring data points.
\par
Using a fixed number of particles per radial bin leads to unequal bin sizes. Since we calculate mean quantities within fixed radial ranges in our further analysis of the profiles, this is a undesired behaviour. Therefore, we apply an additional cubic spline interpolation to the profile, generating a profile with $400$ equidistant sample points.
\par
Demanding a constant number of particles per annuli introduces a natural limit for the radial extent of the kinematic profiles. If there are fewer than $1500$ particles beyond the last considered sampling point, the profile is truncated.

\subsection{Merger Mass Fraction Determination} \label{sec:Merger_Tree}

In Sec. \ref{sec:formation_pathway} we will follow the evolution of individual halos through cosmic time to explore the processes that lead to their $z=0$ properties. Within the $\Lambda$CDM paradigm, structures grow hierarchically through mergers that inevitably affect the galaxies that reside within halos. Thus, merger trees, which comprise information about progenitors, are key to understanding the processes shaping the present-day properties of galaxies. Details about the merger tree construction method used in this work are outlined in App.~\ref{App:merger_tree}.
\par
While merger trees provides a meaningful way to trace the main central structure of a halo, it gives rise to an issue when calculating mass-ratios of two merging objects: At the time of the merger identification $z_{\mathrm{merg}}$, the objects could potentially have experienced significant tidal stripping and other environmental effects, leading to an artificially false estimate of the mass. We correct for this effect by defining the mass of the less massive merger participant $M_{\mathrm{sat}}$ to be its maximum mass before the merger is identified:
\begin{ceqn}
\begin{equation}
M_{\mathrm{Sat}}=\mathrm{max}\left([M_{*}(z):z>z_{\mathrm{merg}}]\right)
\end{equation}
The corresponding mass of the host $M_{\mathrm{Host}}$ is determined at the same moment in time $z_{\mathrm{max}}$:
\begin{equation}
M_{\mathrm{Host}}=M_*(z_{\mathrm{max}})
\end{equation}
\par
Throughout this study we classify a merger event into three merger classes based on the mass-ratio $M_{\mathrm{Sat}}/M_{\mathrm{Host}}$:
\begin{itemize}
	\item \textbf{Mini merger:}
	\begin{equation}
		0.1 > \frac{M_{\mathrm{Sat}}}{M_{\mathrm{Host}}}
	\end{equation}
	\item \textbf{Minor merger:}
	\begin{equation}
		0.3 >\frac{M_{\mathrm{Sat}}}{M_{\mathrm{Host}}} > 0.1
	\end{equation}
	\item \textbf{Major merger:}
	\begin{equation}
		\frac{M_{\mathrm{Sat}}}{M_{\mathrm{Host}}} > 0.3.
	\end{equation}
\end{itemize}
\end{ceqn}
Merger events with $M_{\mathrm{Sat}}/M_{\mathrm{Host}} < 0.01$ are considered to be smooth accretion.

\section{Radial $\lambda$ Profiles out to $5 R_{\mathrm{1/2}}$} \label{sec:radial_profiles_put_to_5Re}

\subsection{Qualitative Connection Between Kinematic Features and the $\lambda(R)$ Profile Shape} \label{sec:cone_kin_prof_shape}

\begin{figure*}
		\centering
        \begin{center}
                \includegraphics[width=0.8\textwidth]{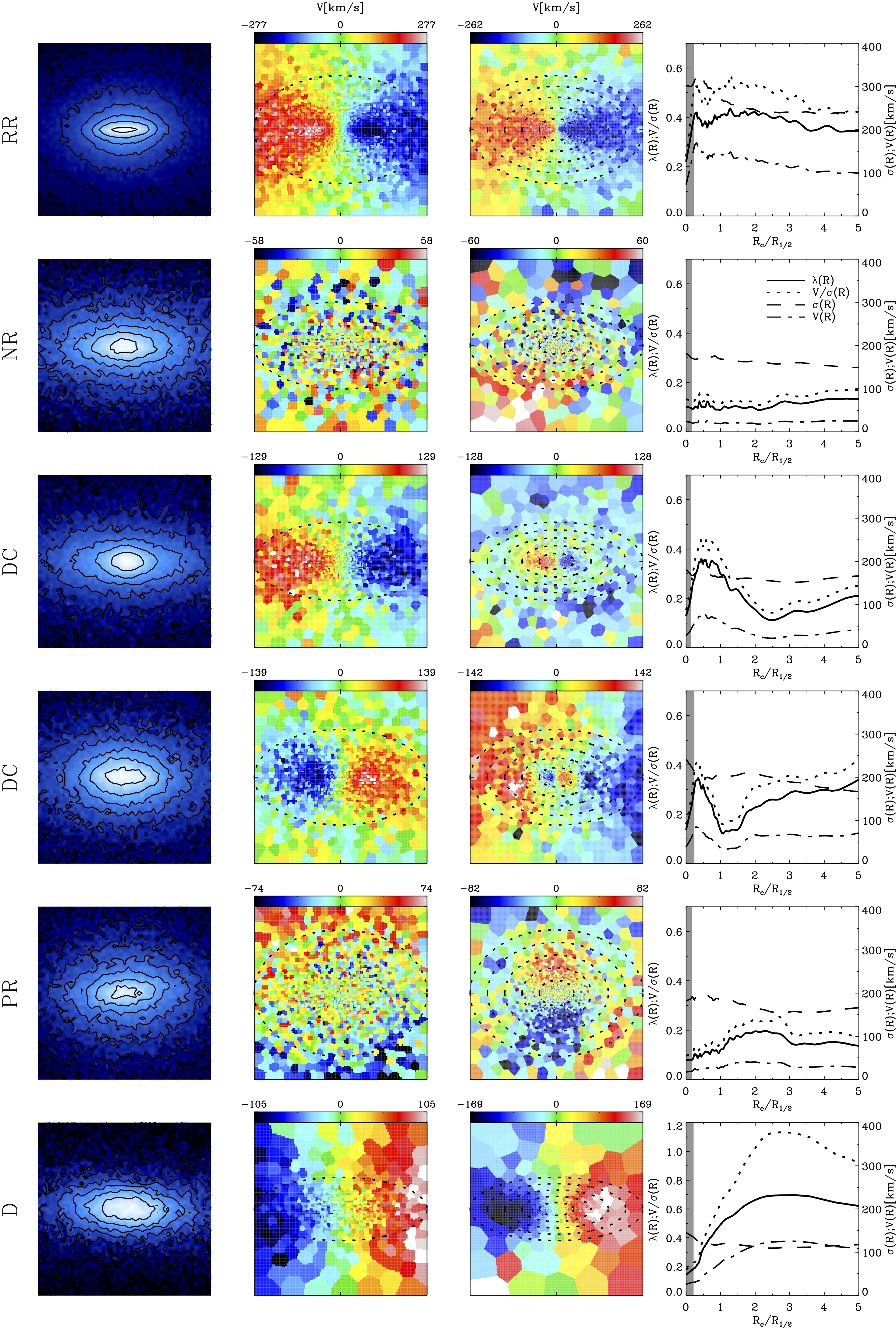}
        \end{center}
        \caption{Each row displays an example of a typical kinematic feature found in the centre of galaxies. From top to bottom: Regular-Rotator, Non-Rotator, Rotating-Core, Rotating-Core, Prolate-Rotator, and Disk. From left to right the panels show the density map, line-of-sight velocity map on a scale of $1R_{\mathrm{1/2}}$, a zoom-out line-of-sight velocity map to $5R_{\mathrm{1/2}}$, and the corresponding radial $\lambda(R)$ (solid), $(V/\sigma)(R)$ (dotted), velocity (dashed dotted), and dispersion (dashed) profile. The dashed black lines in the velocity maps show ellipses with the galaxies' axis ratio and semi-major axis length of integer multiples of $R_{\mathrm{1/2}}$. The grey shaded area in right panels marks the unresolved region below two softening length.}
        {\label{fig:fig_1}}
\end{figure*}

To visualise the connection between the kinematic maps and the $\lambda(R)$ profile, each row in Fig. \ref{fig:fig_1} shows an example of the five kinematic groups defined in \ref{sec:sample_selection}. For the group of DCs we show two examples illustrating the two kinds of kinematic configurations comprised in this group. From left to right the panels display a density map with isophotes, velocity map on a scale of $1R_{\mathrm{1/2}}$, velocity map on a scale of $5R_{\mathrm{1/2}}$, and the corresponding $\lambda(R)$ (solid), $V/\sigma(R)$ (dotted), $\sigma(R)$ (dashed), and $V(R)$ (dashed-dotted) radial profiles. The dashed black ellipses in the velocity maps mark isophotes with semi-major axis length of integer multiples of $R_{\mathrm{1/2}}$.
\par
The RR (first row) shows the characteristic velocity pattern of a centrally fast rotator. Accordingly, this galaxy has $\lambda_{\mathrm{R_{1/2}}}=0.42$ clearly in the fast rotating regime. The regular rotation pattern extends beyond $1R_{\mathrm{1/2}}$ with a mild decrease in the velocity amplitude. This behaviour is reflected in the $\lambda(R)$ profile: it increases steeply in the very centre, followed by an almost constant section with a mild decrease beyond $2R_{\mathrm{1/2}}$. At $\sim 5R_{\mathrm{1/2}}$, the profile reaches its minimum of $\sim 0.35$ still, in the fast rotating regime.
\par
The NR (second row) represents a typical example of a non-rotating slow rotator based on its velocity field in the centre as well as based on its stellar spin $\lambda_{\mathrm{R_{1/2}}}=0.07$. The profile exhibits only minor variations out to $3R_{\mathrm{1/2}}$, and a moderate increase beyond this radius out to the maximal probed radius. With a maximum of $\sim 0.15$ the galaxy exhibits low rotational support in the entire investigated radial range.
\par
The first DC (third row) shows significant variations in the velocity map and the $\lambda(R)$ profile: this galaxy exhibits an apparent rotating core with a non-rotating surrounding stellar halo. Already towards the centre, there is an apparent drop in the amplitude of the velocity. Furthermore, the regular rotating velocity pattern disperses within $2R_{\mathrm{1/2}}$. Beyond this radius there is only minor evidence of low level rotation. The appearance of the velocity maps is consistent with a rotating disk embedded in a non-rotating stellar halo. The $\lambda(R)$ profile shows a peak in the centre followed by a drastic decrease out to $2.5R_{\mathrm{1/2}}$, and a subsequent mild incline capturing the transition from the rotating core to the non-rotating surroundings.
\par
The second DC (fourth row) shows a rotating core component that is embedded in a halo that rotates with a kinematic position angle close to $180\degree$ with respect to the centre. Accordingly, $\lambda(R)$ has a significant depression exactly at the transition radius between the two kinematic subcomponents. Subsequently, the profile plateaus at $3R_{\mathrm{1/2}}$ followed by a moderate increase.
\par
The fifth row displays a member of the special group of PRs. These galaxies show ordered rotation around the morphological major axis. Within $1R_{\mathrm{1/2}}$, the velocity map resembles that of a slow rotator. In the radial range of $1-2R_{\mathrm{1/2}}$, the prolate rotation is apparent. Beyond $3R_{\mathrm{1/2}}$, the velocity pattern becomes asymmetric: While the rotation in the upper part vanishes completely, the rotation in the lower part of the map extends significantly further out. The asymmetric lower part might be caused by stripped particles of an in-falling galaxy which caused the prolate rotation signal. Within the Illustris simulation \citet{2017ApJ...850..144E} found that prolate rotation is strongly correlated with the last significant merger a galaxy experienced. Furthermoe, slight asymmetries in the velocity map during the relaxation process after the merger, as found in our example, are possible to occur. However, investigating this in more detail is beyond the scope of this paper.
\par
As expected, the D in the last row shows flatter isophotes than the other examples. The ordered motion is apparent out to a scale of $5R_{\mathrm{1/2}}$. The $\lambda(R)$ profile reaches up to values of $\approx 0.7$ and therefore the extremely rotationally supported range. This is mainly due to the low $\sigma(R)$ in comparison to the RR example. This is the only example where we find a crossing of of $V(R)$ and $\sigma(R)$ at $\approx 2R_{\mathrm{1/2}}$, implying clear rotational support.
\par
Comparing the RCs and RR clearly shows that galaxies that exhibit a similar velocity feature and $\lambda(R)$ profile within $1R_{\mathrm{1/2}}$ can have very different outer halo kinematics. Hence, it is not sufficient to solely investigate the central $1R_{\mathrm{1/2}}$ to assess the full kinematic state of a galaxy. Furthermore, these variations in the halo kinematics are connected to the distinct formation pathway of a galaxy and therefore encode valuable information about the processes that shape the galaxy. 
\par
The $(V/\sigma)(R)$ profile shows the same behaviour as $\lambda(R)$ in every aspect and thus can also be used to trace the radial behaviour (see also App.~\ref{AppA}).

\subsection{Quantifying Profile Shape}

\subsubsection{Profile Gradients at Fixed Radii and Maximum Amplitude} \label{sec:comparison_observations}

We compare the shape of the $\lambda(R)$ profiles in the simulation to recent observations from the SLUGGS \citep{2017MNRAS.467.4540B}, SAMI \citep{2018MNRAS.480.3105F}, and ePN.S \citep{2018A&A...618A..94P} survey. Due to the challenging task of observing spectra at low surface brightnesses in the stellar halo, the three observational comparison samples use different methods to determine velocity maps out to large radii. \citet{2018MNRAS.480.3105F} uses direct integral-field observations with a radial coverage out to typically $\sim 1.5$--$2.5 R_{\mathrm{e}}$. \citet{2017MNRAS.467.4540B} utilises the SKiMS technique which uses DEIMOS slit observations to capture the underlying stellar velocity field and apply a novel interpolation method to obtain a continuous velocity map with a radial coverage out to $\sim 2$--$3R_{\mathrm{e}}$. A similar approach is used by \citet{2018A&A...618A..94P}, however using planetary nebulae as discrete tracers, which allow for a significantly larger radial coverage out to typically $\sim 6 R_{\mathrm{e}}$.
\par
\begin{figure}
        \begin{center}
                \includegraphics[width=0.45\textwidth]{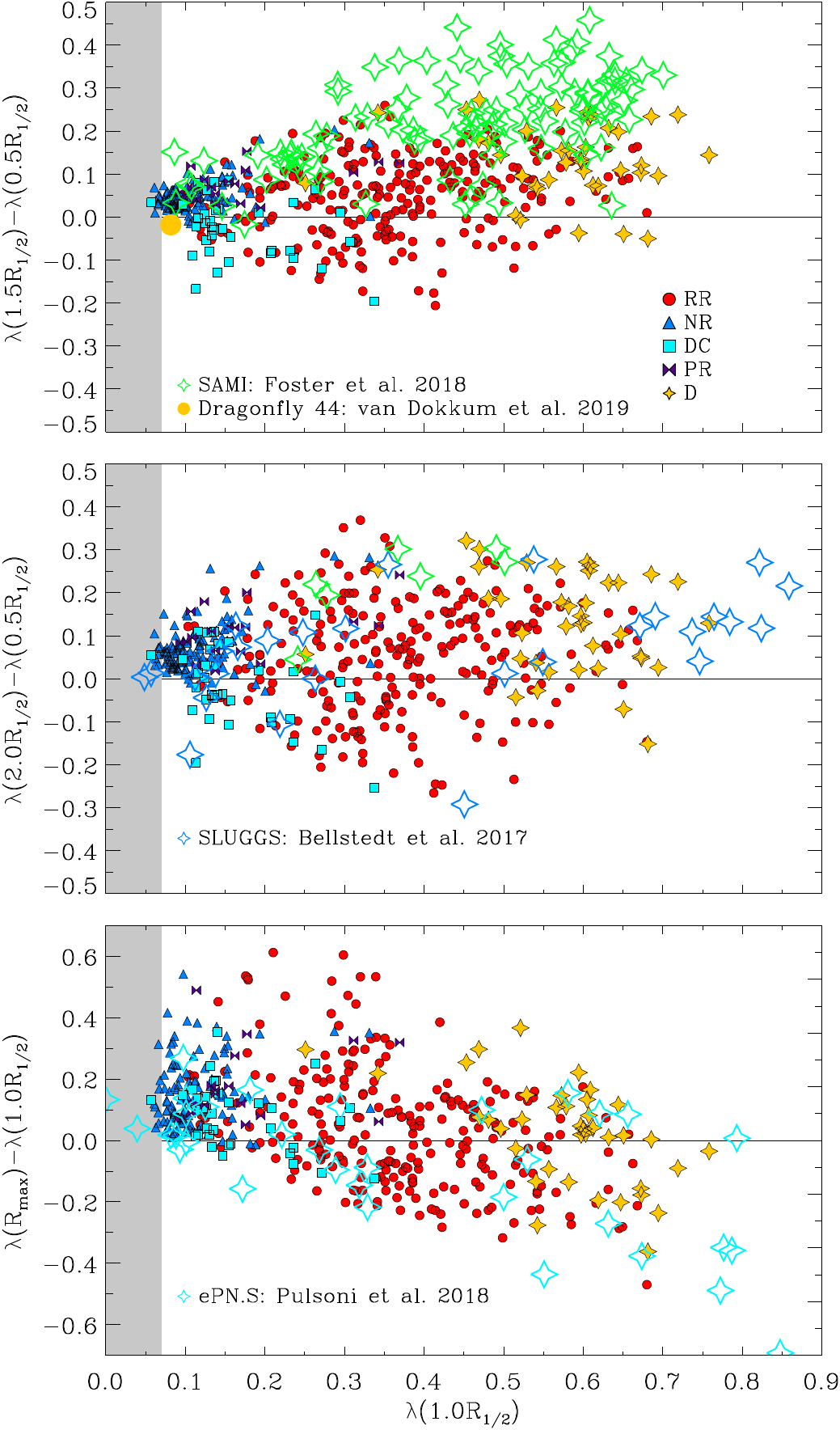}
        \end{center}
        \caption{\textit{Upper Panel:} Gradient in $\lambda(R)$ profile between $1.5R_{\mathrm{1/2}}$ and $0.5R_{\mathrm{1/2}}$ versus the local $\lambda(R)$ measured at $1.0R_{\mathrm{1/2}}$. The galaxies are subdivided according to their central kinematic group Regular-Rotators (red circles), Non-Rotators (blue triangles), Distinct-Cores (cyan square), Prolate-Rotators (lilac bowtie), and Disks (yellow stars). The grey shaded region marks the due to resolution unreachable region. Green symbols show observational results from the SAMI survey \citep{2018MNRAS.480.3105F}, and the orange symbol marks Dragonfly 44 \citep{2019ApJ...880...91V}. \textit{Middle Panel:} Same as upper panel, however the gradient is calculated between $2.0R_{\mathrm{1/2}}$ and $0.5R_{\mathrm{1/2}}$. We include observations from the SAMI survey (green) and the SLUGGS survey \citep[blue,][]{2017MNRAS.467.4540B}. \textit{Lower Panel:} Same as the upper two panels with the gradient calculated between the outermost data point $R_{\mathrm{max}}$ and $1.0R_{\mathrm{1/2}}$. The cyan symbols represent observations from the ePN.S survey extracted from \citet{2018A&A...618A..94P}.}
        {\label{fig:fig_2}}
\end{figure}
Fig. \ref{fig:fig_2} shows the two point gradient of the $\lambda(R)$ profile measured at different points versus the local $\lambda(1 R_{\mathrm{1/2}})$ for our sample, split up into the kinematic groups, in comparison to the above mentioned observations. The upper panel shows $\lambda(1.5 R_{\mathrm{1/2}})-\lambda(0.5 R_{\mathrm{1/2}})$ vs. $\lambda(1 R_{\mathrm{1/2}})$ in comparison to SAMI, while the middle panel shows $\lambda(2.0 R_{\mathrm{1/2}})-\lambda(0.5 R_{\mathrm{1/2}})$ vs. $\lambda(1 R_{\mathrm{1/2}})$ in comparison to SAMI and SLUGGS. The lower panel shows the $\lambda(R_{max})-\lambda(1.0 R_{\mathrm{1/2}})$ vs. $\lambda(1 R_{\mathrm{1/2}})$ in direct comparison to ePN.S. The SLUGGS sample contains $15$ S0 galaxies and $13$ elliptical galaxies, while the SAMI survey provides $\lambda(2.0 R_{\mathrm{1/2}})-\lambda(0.5 R_{\mathrm{1/2}})$ values for $7$ galaxies where we do not have access to a morphological classification. In addition, we have $\lambda(1.5 R_{\mathrm{1/2}})-\lambda(0.5 R_{\mathrm{1/2}})$ values for $107$ galaxies of mixed morphologies, including late-type galaxies, from the SAMI survey. From the ePN.S survey we include $33$ ETGs. Here, we have access to $\lambda(R_{max})-\lambda(1.0 R_{\mathrm{1/2}})$, where $R_{max}$ is in the ranges $3.0 R_{\mathrm{1/2}}$-$13 R_{\mathrm{1/2}}$. We also derive a $\lambda(R)$ profile for the ultra diffuse galaxy Dragonfly 44 from the data provided in \citet{2019ApJ...880...91V}. For the simulation, we calculate the gradients always in the same radial range as the comparison observations.
\par
The middle and lower panels show a good agreement between the simulated sample and the observations. The \textit{Magneticum} galaxies cover the same range of gradients $[-0.3$-$0.35]$ as the observations. Minor discrepancies are visible in the high $1.0R_{\mathrm{1/2}}$ range where the SLUGGS survey reaches values of $0.85$ which are not found in the simulated sample. We suspect, that this might me due to the still to high dispersion in disk galaxies formed in the simulation. More significant disagreements are apparent in the upper panel: The SAMI galaxies exhibit significantly larger gradients than the simulated galaxies. This is even more surprising considering that SAMI does not favour the edge-on projection in the sample selection, which would maximize the gradient, like we are. Furthermore, there is only one object with negative gradient. The low number of objects with negative gradients with respect to the SLUGGS sample was already stated in \citet{2018MNRAS.480.3105F} and quantitatively resolved by accounting for differences in the sample selection and observational effects. One side note, the ultra diffuse galaxy Dragonfly 44 (orange symbol) is a regular galaxy in the plane exhibiting low rotation in the centre and only very minor variations in the $\lambda(R)$ profile.
\par
For the simulated sample, the overall distribution of kinematic groups is as expected in all panels: The RRs and Ds show the largest gradients, while the NR and PRs predominately populate the small gradient regime below $\sim 0.1$. Furthermore, most of the DCs have negative gradients, reflecting the drop in the rotational support seen in the velocity map. Interestingly, the gradient for RRs and Ds shift significantly to negative values when the outer sampling point is shifted to larger radii in agreement with the observations. This reflects that many of these object have a centrally increasing profile ($<R_{\mathrm{1/2}}$) that decreases at larger radii. 
\begin{figure}
        \begin{center}
                \includegraphics[width=0.48\textwidth]{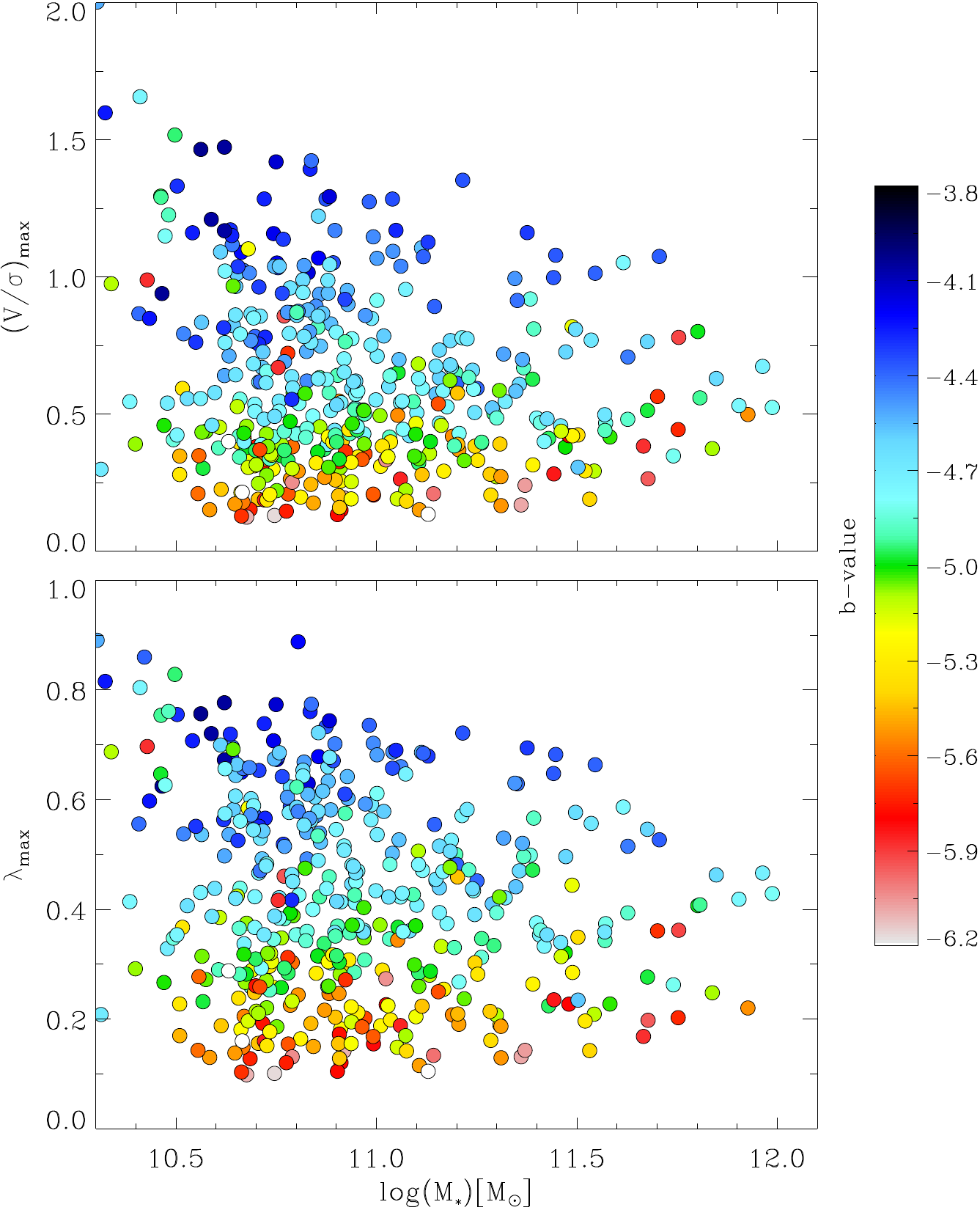}
        \end{center}
        \caption{The maximum of the $(V/\sigma)(R)$ profile (\textit{upper panel}) and the $\lambda(R)$ profile (\textit{lower panel}) as function of stellar mass. Data points are colour-coded according to their b-value as an proxy for their morphology. A b-value larger than $-4.3$ is characteristic for disc galaxies, while the morphology transitions to earlier types with decreasing b-value.}
        {\label{fig:fig_5}}
\end{figure}
\par
Apart from the radial variations in the profile, the maximum amplitude is a characteristic quantity describing radial profiles. Therefore, Fig. \ref{fig:fig_5} shows the correlation between the stellar mass and the maximum amplitude of the $(V/\sigma)(R)$ profiles (upper panel) and the $\lambda(R)$ profiles (lower panel), denoted as $(V/\sigma)_{\mathrm{max}}$ and $\lambda_{\mathrm{max}}$, respectively. The symbols are coloured according to the b-value as a proxy for morphology. It represents the direct comparison with Figure 4 of \citet{2018MNRAS.480.3105F}. Although Fig. \ref{fig:fig_5} covers a larger mass range than \citet{2018MNRAS.480.3105F}, we confirm the trend that there is no overall correlation between the profile maximum and the stellar mass. This holds true for the directly comparable $(V/\sigma)_{\mathrm{max}}$ as well as for $\lambda_{\mathrm{max}}$. However, a clear correlation between the b-value, and therefore the morphology, and the location in this plane is visible: There is a continuous sequence of decreasing b-values with decreasing $(V/\sigma)_{\mathrm{max}}$ and $\lambda_{\mathrm{max}}$, which reflects the higher rotational support of late-type galaxies. It is, however, difficult to judge whether galaxies of equal b-value lie on relations with varying slope as found by \citet{2018MNRAS.480.3105F} for different morphological types, given the large scatter.

\subsubsection{Refined Classification Scheme}
\label{sec:refined_classification_scheme}
\begin{figure*}
       \begin{center}
                \includegraphics[width=0.89\textwidth]{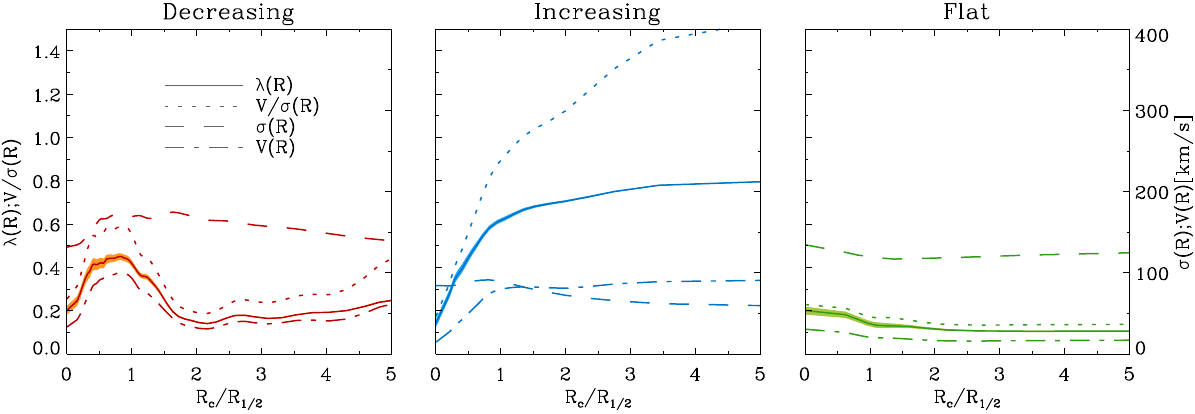}
        \end{center}
        \caption{Hand-picked examples for a decreasing (\textit{left panel}), increasing (\textit{middle panel}) and flat profile (\textit{right panel}). In each panel the different lines distinguish $\lambda(R)$ (solid), $V/\sigma(R)$ (dotted), $\sigma(R)$ (dashed) and $V(R)$ (dashed dotted). The shaded area around the $\lambda(R)$ profiles marks the $16\mathrm{th}$ and $84\mathrm{th}$ percentile of a bootstrapping.}
       {\label{fig:fig_0}}
\end{figure*}
As discussed in Sec.~\ref{sec:cone_kin_prof_shape}, galaxies can show significantly different $\lambda(R)$ profile shapes. While the examples shown in Fig. \ref{fig:fig_1} represent single example cases, the aim of this study is to analyse our sample of galaxies in a statistical manner. In order to find statistical meaningful trends, it is necessary to quantify the shapes of the radial $\lambda(R)$ profiles and, if possible, to classify them.
\par
A comprehensive visual inspection of our sample revealed three characteristic shapes:
\begin{itemize}
	\item Decreasing Profile: Galaxies showing a central peak and a subsequent decrease in the outer region. The position of the peak ranges from $\sim 0.5R_{\mathrm{1/2}}$ to $\sim 2.0R_{\mathrm{1/2}}$.
	\item Increasing Profile: Galaxies showing a continuous increase out to the maximally probed radius. The increase is either close to linear with varying slope or shows a steep increase in the inner region followed by a plateau at larger radii.
	\item Flat Profile: Galaxies showing only minor variations over the entire radial range. The amplitude of the profile ranges from low $\lambda$ values in the slow rotating regime to high fast rotating values, albeit the former case is much more frequent.
\end{itemize}
\par
Earlier observational studies use the gradient of the $\lambda(R)$ or $V/\sigma(R)$ profile calculated from two sampling points in the inner region ($\sim 0.5R_{\mathrm{1/2}}$) and outer region ($\sim 1.5-2R_{\mathrm{1/2}}$) depending on the covered radial range \citep{2014ApJ...791...80A,2017MNRAS.467.4540B,2018MNRAS.480.3105F,2018A&A...618A..94P} as discussed in Sec. \ref{sec:comparison_observations}.
\par
Fig. \ref{fig:fig_0} shows an example for each of the three classes. From left to right: Decreaser, Increaser and Flat. In each panel the various lines display the $\lambda(\mathrm{R})$ (solid), $V/\sigma(R)$ (dotted), $\sigma(R)$ (dashed) and $V(R)$ (dashed-dotted) profiles. The shaded areas mark the $1\sigma$confidence interval derived from $500$ bootstraps within each radial bin, demonstrating the minor statistical noise in the profiles. While the last two categories can be captured by a simple gradient with two sampling points, the varying position of the peak for the decreasing profiles makes this simple approach unsuitable.
\par
Visual inspection and of the profiles showed that the following classification disentangles the three characteristic shapes:
\begin{itemize}
	\item \textit{Decreasing:}
	\begin{ceqn}
	\begin{equation}
		\overline{\lambda \bigl (2.0<R<3.5 \bigr)}-\overline{\lambda \bigl (0.5<R<2.0 \bigr)} < -0.04
		\label{eqn:gradient_classification}
	\end{equation}
	\end{ceqn}
	\item \textit{Increasing:}
	\begin{ceqn}
	\begin{equation}
		\overline{\lambda \bigl (2.0<R<3.5 \bigr)}-\overline{\lambda \bigl (0.5<R<2.0 \bigr)} > 0.04
	\end{equation}
	\end{ceqn}
	\item \textit{Flat:}
	\begin{ceqn}
	\begin{equation}
	\begin{aligned}
		|\overline{\lambda \bigl (0.5<R<R_{\mathrm{max}} \bigr)} - \lambda_{\mathrm{max}}| < 0.09 ~~\& \\
		|\overline{\lambda \bigl (0.5<R<R_{\mathrm{max}} \bigr)} - \lambda_{\mathrm{min}}| < 0.09
	\end{aligned}
	\end{equation}
	\end{ceqn}
\end{itemize}
$\lambda_{\mathrm{max}}$ and $\lambda_{\mathrm{min}}$ are the maximum and minimum values reached by $\lambda(R)$, respectively. Throughout this paper we will use this classification, correlate it with various galaxy properties, and investigate its connection to the formation history of the galaxy.
\par
In total, $84\%$ of the sample can be assigned to one of the defined classes. Accordingly, $16\%$ are unclassified. These galaxies typically show large variation in their profile with several maxima and minima, and are tidally interacting or in the process of merging.
\par
Increasing profiles comprise $47\%$ of our total sample and therefore represent the most frequent profile type. Decreasing and flat profiles are equally abundant with $19\%$ and $18\%$, respectively. The predominance of increasing profiles is in qualitative agreement with former observational studies by \citet{2017MNRAS.467.4540B}, \citet{2018MNRAS.480.3105F} and \citet{2014ApJ...791...80A}. While these studies find disagreements in the actual numbers, which might be due to small sample sizes and selection biases, they consistently find a significantly larger fraction of increasing profiles than other shapes. As shown in App.~\ref{AppA} this also holds when using the $V/\sigma$ instead of $\lambda$.


\section{Profile Shape Correlation with Galaxy Properties} \label{sec:Profile_Shape_Correlation_with_Galaxy_Properties}

To examine the imprint of the large-scale stellar kinematics on other fundamental galaxy properties, we investigate the connection between the $\lambda(R)$ profile shape and the central stellar kinematics, the stellar mass, and the morphology via the b-value. Several earlier studies have addressed this topic in the past: Having access to a large sample of $384$ galaxies, \citet{2018MNRAS.480.3105F} found that the radial dynamical support is linked to the visual morphology, however with a significant intrinsic scatter. Furthermore, \citet{2017MNRAS.467.4540B} found a strong separation between elliptical galaxies and S0 galaxies in the plane of stellar spin gradient and local stellar spin at $1\mathrm{R_{1/2}}$ giving more physical meaning to the morphological distinction.

\subsection{Correlation with Central Stellar Kinematics}

First we investigate the connection between profile shape and the notion of centrally fast and slow rotators. To distinguish between fast and slow rotators, we use the quantity $\lambda_{\mathrm{F/S}} = \lambda_{\mathrm{R_{1/2}}} / 0.31 \sqrt{\epsilon}$, with fast rotators having $(\lambda_{\mathrm{F/S}}>1)$ and slow rotators having $(\lambda_{\mathrm{F/S}}<1)$, according to the criterion introduced by \citet{2011MNRAS.414..888E}.
\par
\begin{figure}
        \begin{center}
                \includegraphics[width=0.45\textwidth]{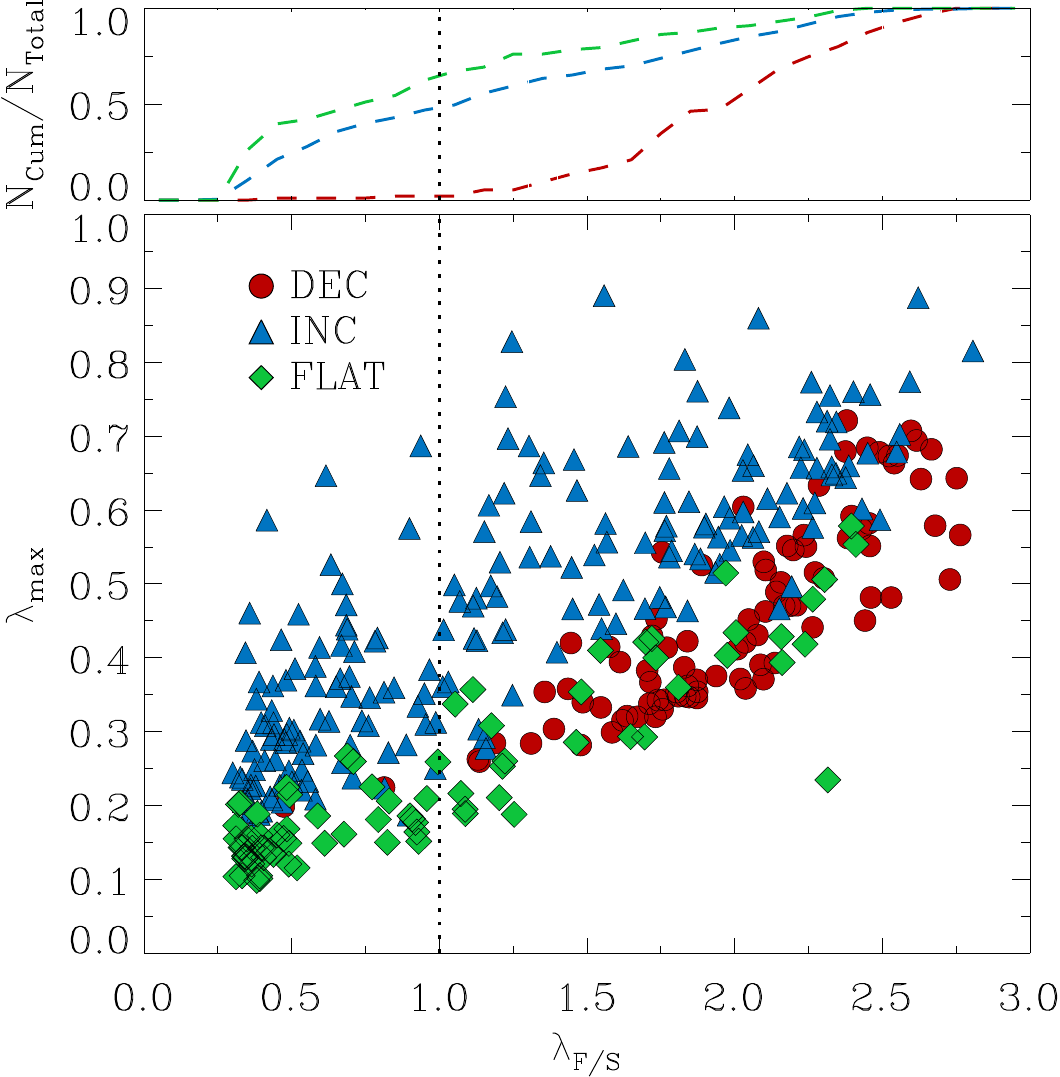}
        \end{center}
        \caption{\textit{Lower panel:} the quantity $\lambda_{\mathrm{F/S}}=\lambda_{\mathrm{R_{1/2}}} / 0.31 \sqrt{\epsilon}$ versus the maximal amplitude of the $\lambda(R)$ profile $\lambda_{\mathrm{max}}$, split up into decreasing (red circles), increasing (blue triangles) and flat (green squares) profiles. For central fast rotators, $\lambda_{\mathrm{F/S}}$ is larger than one, accordingly $\lambda_{\mathrm{F/S}}$ is smaller than one for slow rotators. \textit{Upper panel:} the cumulative number of galaxies ($\mathrm{N_{Cum}}$) normalised by the total number of galaxies ($\mathrm{N_{Total}}$) for each class.}
        {\label{fig:fig_m1}}
\end{figure}

The main panel of Fig. \ref{fig:fig_m1} relates this quantity to $\lambda_{\mathrm{max}}$, which is the maximal amplitude of the $\lambda(R)$ profile, for increaser (triangles), decreaser (circles), and flats (diamonds). The top panel displays the cumulative distribution of $\lambda_{\mathrm{F/S}}$ for each class normalised to the number of members of the respective class.
\par
The increasers cover the full range of measured $\lambda_{\mathrm{F/S}}$, and therefore contain both fast and slow rotators. Approximately $50\%$ are classified as fast rotators, and $50\%$ as slow rotators. In general, the increasers are separated from the other two classes having larger $\lambda_{\mathrm{max}}$ values at all $\lambda_{\mathrm{F/S}}$. Only in the high $\lambda_{\mathrm{F/S}}$ regime we find a minor overlap with the decreasing class. The maximum $\lambda_{\mathrm{max}}$ reached by increaseres is $0.9$, which is in the extremely rotational dominated domain. The distribution features a well defined lower envelop: with increasing $\lambda_{\mathrm{F/S}}$ the minimum $\lambda_{\mathrm{max}}$ increases accordingly. For a given $\lambda_{\mathrm{F/S}}$, the increasing profiles show a significantly larger vertical scatter than the two other classes.
\par
Similar to the increasers, the flats cover a wide range of $\lambda_{\mathrm{F/S}}$, and hence comprise slow and fast rotators. However, the concentration of the distribution is located in the slow rotating regime. Approximately $70\%$ are centrally slow rotating, while $30\%$ are fast rotating. It might be that the fast rotating flats are actually increasers, but low mass and therefore the steep inner rise of the profile falls below the resolution limit. Those objects show a constant rotational support over the whole range of investigated radii. 
\par
The overall distribution of flats is well constrained by an upper and lower envelope following a very similar relation, describing a fairly tight connection: as $\lambda_{\mathrm{F/S}}$ increases, $\lambda_{\mathrm{max}}$ increases as well. This class also contains an extreme outlier with a $\lambda_{\mathrm{max}}=0.24$ and a anomalous low $\lambda_{\mathrm{F/S}}$. We checked this object and found that it is unusually round, which causes $\lambda_{\mathrm{F/S}}$ to be small. This particular object is an outlier in every correlation investigated in \citet{2018MNRAS.480.4636S}. However, looking into this in detail is beyond the scope of this paper.
\par
The decreasing profiles populate the high $\lambda_{\mathrm{F/S}}$ regime separated from the increasers by generally lower $\lambda_{\mathrm{max}}$ values. Interestingly, except for one object, all of the decreasers are classified as central fast rotators. This supports the findings of \citet{2014ApJ...791...80A} and \citet{2018MNRAS.480.3105F} that these profiles resemble stellar rotating structures embedded in a non-rotating halo. As aforementioned, we find a minor overlap between decreasers and increasers at high $\lambda_{\mathrm{F/S}}$ values. Similar to the flats the decreasers exhibit a rather tight correlation that increases with increasing $\lambda_{\mathrm{F/S}}$. This trend is expected since decreasing profiles reach their $\lambda_{\mathrm{F/S}}$ at low radii close to where $\lambda_{\mathrm{F/S}}$ is measured.
\par
Overall, the three classes populate distinctly different regions in this plane with only minor overlaps. Comparing the flat and decreasing profiles, it seems like the decreasers represent a natural extension of the relation found for flats to larger $\lambda_{\mathrm{F/S}}$. We suspect that there is an evolution channel from decreasers to flats: decreasers seem to evolve towards flat profiles due to a less significant peak in the centre. We therefore suspect the decreasing and flat profiles to have a similar merger history however the flat profiles experience at least one high mass fraction merger that kills off the rotation in the centre and transform an decreaser into a flat.
\par
In order to investigate how the profile shape is connected to a refined classification of the central kinematics, we apply the central kinematic classification introduced in Sec. \ref{sec:cone_kin_prof_shape}. Tab. \ref{tab:kin_group} summarizes the fraction of decreasing, increasing and flat profiles for each kinematic group.
\par
The RRs comprise objects from all three profile groups. As expected, the most frequent group within the RRs are the increasers with $49\%$. The decreaseres encompass $36\%$ of the RR, while $15\%$ are flat. The flat profiles in this group correspond to the fast rotating flats in Fig. \ref{fig:fig_m1} exhibiting high amplitudes in the profile. 
\begin{table}
	\centering
	\caption{The statistical distribution of decreasing, increasing and flat profiles within the five kinematic groups.}
	\label{tab:kin_group}
	\begin{tabular}{ccccc}
		\hline
		 & Decreasing & Increasing & Flat & $\mathrm{N_{gal}}$\\
		\hline
		\hline
		RR & 36\% & 49\% & 15\% & 208\\
		NR & 1\%& 61\%& 38\% & 101\\
		DC & 15\%& 48\%& 37\% & 27\\
		PR & 0\%& 100\%& 0\% & 14\\
		D & 37\%& 63\%& 0\% & 30\\
		\hline
	\end{tabular}
\end{table}
\par
As expected, the NRs are in general less rotationally support than the RRs. The fraction of flat profiles is $38\%$ and therefore twice as frequent as for the RRs. Therefore, objects with low rotational support in the centre stay relatively often pressure supported out to $5R_{\mathrm{1/2}}$. This reflects the low number of fast rotating flats found in Fig. \ref{fig:fig_m1}. Interestingly, the NRs include a large fraction of increasers of $61\%$ even higher than the average. However, the shape of the increasing profiles of the NRs differs from the RRs: The amplitude is in general smaller. Furthermore, the increase is more modest and closer to linear.
\par
An interesting feature of the DCs is the fact, that they comprise a similar fraction of flats ($37\%$) and increasers ($48\%$), and only a small fraction of decreasers ($15\%$). Due to the definition of DCs one would expect to preferentially find decreasing profiles in this kinematic group, since a decrease of the rotation in the LOSV-map is characteristic for this group, except for the members with counter rotation which are however extremely rare ($5$ galaxies). Hence, the visual appearance of the LOSV-map within the central $1R_{\mathrm{1/2}}$ is not always sufficient to deduce the profile shape at larger radii demonstrating the complex connection between central and halo kinematics. 
\par
All of the $14$ PRs have an increasing $\lambda(R)$ profile. This signal shows that for PRs the central and halo kinematics are coupled and are determined by a process that affects the centre and the halo. This is in line with the findings of \citet{2017ApJ...850..144E} that a significant merger sets the prolate rotation which can alter the orbital configuration in the centre and the halo.
\par
A noticeable feature in the D group is the occurrence of increasing and decreasing $\lambda(R)$ and $(V/\sigma)(R)$ profiles. This means, that extremely rotational support disks can exist within a rotating ($63\%$ increaser) and non-rotating ($37\%$ decreaser) halo, i.e. 1/3 in a non-rotating halo. Therefore, the process causing the pressure support in the halo, and hence the difference in the formation of the halo, does not effect the very centre of the galaxy.

\subsection{Correlation with Stellar Mass}

Earlier studies found a strong correlation between the stellar mass of a galaxy and its central kinematics. According to these studies the fraction of slow rotators increases significantly with stellar mass \citep{2017ApJ...844...59B,2013ApJ...778..171J,2017MNRAS.471.1428V}. In order to extend this to the halo kinematics, we study the relation between the profile shape and stellar mass.
\par
\begin{figure}
        \begin{center}
                \includegraphics[width=0.48\textwidth]{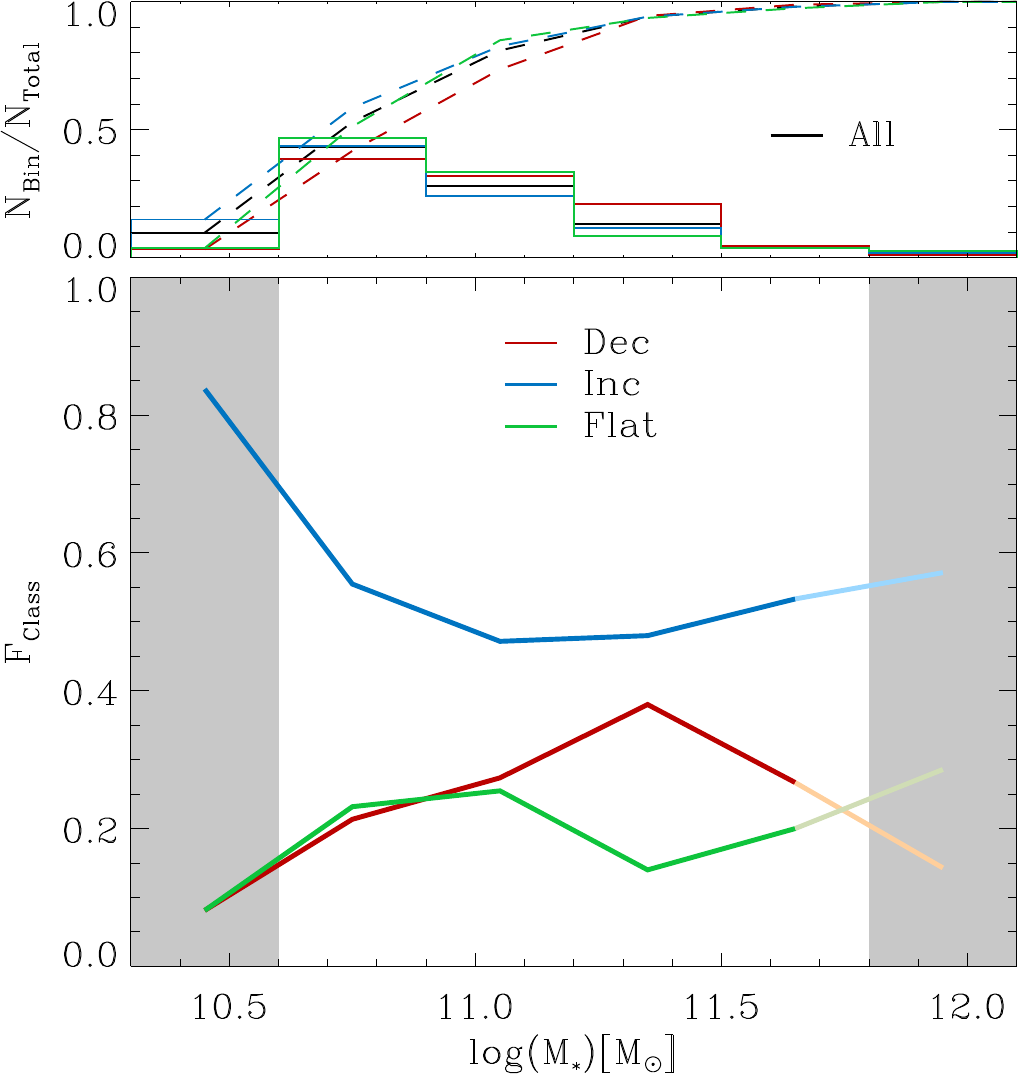}
        \end{center}
        \caption{The \textit{lower panel} depicts the connection between the fraction $\mathrm{F_{Class}}$ of increasing (blue), decreasing (red) and flat (green) profiles and the stellar mass normalised within each mass-bin. The grey shaded region marks the, due to resolution and box-size, unreliable mass-range. Bins with less than $10$ galaxies are shown in lighter colours. The histogram in the \textit{upper panel} shows the absolute number of galaxies in each mass bin ($\mathrm{N_{Bin}}$)normalised by the total number of galaxies ($\mathrm{N_{Total}}$) of the respective class, while the dashed line represents the according cumulative distribution. The black curves represent the all galaxies in the sample.}
        {\label{fig:fig_3}}
\end{figure}

Fig. \ref{fig:fig_3} shows the stellar mass distribution of the three profile classes in $6$ bins. To investigate the distribution within each bin the lower panel depicts the fraction of increasing (blue), decreasing (red) and flat (green) profiles normalised within each mass-bin. In contrast, the upper panel shows the distribution of the total number of galaxies ($\mathrm{N_{Bin}}$) normalised to the total number of galaxies $\mathrm{N_{Total}}$ (solid line), and the cumulative fraction (dashed line) in each individual class. We emphasize that, due to the limited box size of the simulation, our sample is not complete in the high mass regime.
\par
Throughout the whole probed mass range, the increasers are the most frequent class. In the lowest mass bin, the increaseres represent $87\%$ of the total sample. For higher masses, the fraction of increasing profiles drops to $\approx 50\%$ and shows no further mass dependence. This is a first hint towards a connection between the profile shape and the morphology of galaxies, since we expect the largest fraction of rotational supported late-type and S0 galaxies in the low mass bin.
\par
The fraction of decreaser and flats rises concurrently from $\approx 7\%$ at $\mathrm{log(M_*/M_{\odot})=10.5}$ up to $\approx 25\%$ at $\mathrm{log(M_*/M_{\odot})=11.1}$. Beyond $\mathrm{log(M_*/M_{\odot})=11.1}$, the decreasers become more frequent, rising up to $40\%$ and then gradually dropping back to $20\%$. The flats show a contrasting behaviour with a minor degression and a subsequent rise. Therefore, the relative fraction of decreasing and flat profiles stays constant within the considered mass range.
\par
Investigating the normalised distribution for each individual class in the upper panel of Fig. \ref{fig:fig_3}, we find a similar behaviour for the three classes: All distributions peak in the mass-bin centred on $\mathrm{log(M_*/M_{\odot})=10.75}$ and show a gradual decrease towards higher masses. The flattening of the cumulative distributions above $\mathrm{log(M_*)=11.5}$ is mainly driven by the lack of high mass objects due to the limited size of the simulated box. According to conclusions drawn from the distribution in the lower panel increasers exhibit the largest excess with respect to the other groups in the lowest mass bin. Overall, we do not find a significant difference in the individual mass dependence among the three classes. In line with our findings, using $V/\sigma(R)$ profiles, which are closely related to $\lambda(R)$, \citet{2018MNRAS.480.3105F} found only a weak trend for more massive galaxies to have slightly larger $V/\sigma(1.5R_{\mathrm{e}})-V/\sigma(0.5R_{\mathrm{e}})$ gradients.

\subsection{Correlation with Morphology via the b-value} \label{sec:cor_morph}

A connection between the $V/\sigma$ gradient with morphology was reported for the SLUGGS survey by \citet{2014ApJ...791...80A} and \citet{2017MNRAS.467.4540B}, and especially by \citet{2018MNRAS.480.3105F} for the SAMI survey. According to these studies the gradients in the $V/\sigma$ profiles get larger when going from early-type to late-type galaxies. Furthermore, \citet{2018MNRAS.480.3105F} found a clear morphology trend in the $(V/\sigma)_{\mathrm{max}}$-$\mathrm{log(M_*)}$ plane that can be understood by the Faber-Jackson \citep{1976ApJ...204..668F} and Tully-Fisher relation \citep{1977A&A....54..661T}.
\par
\begin{figure}
        \begin{center}
                \includegraphics[width=0.48\textwidth]{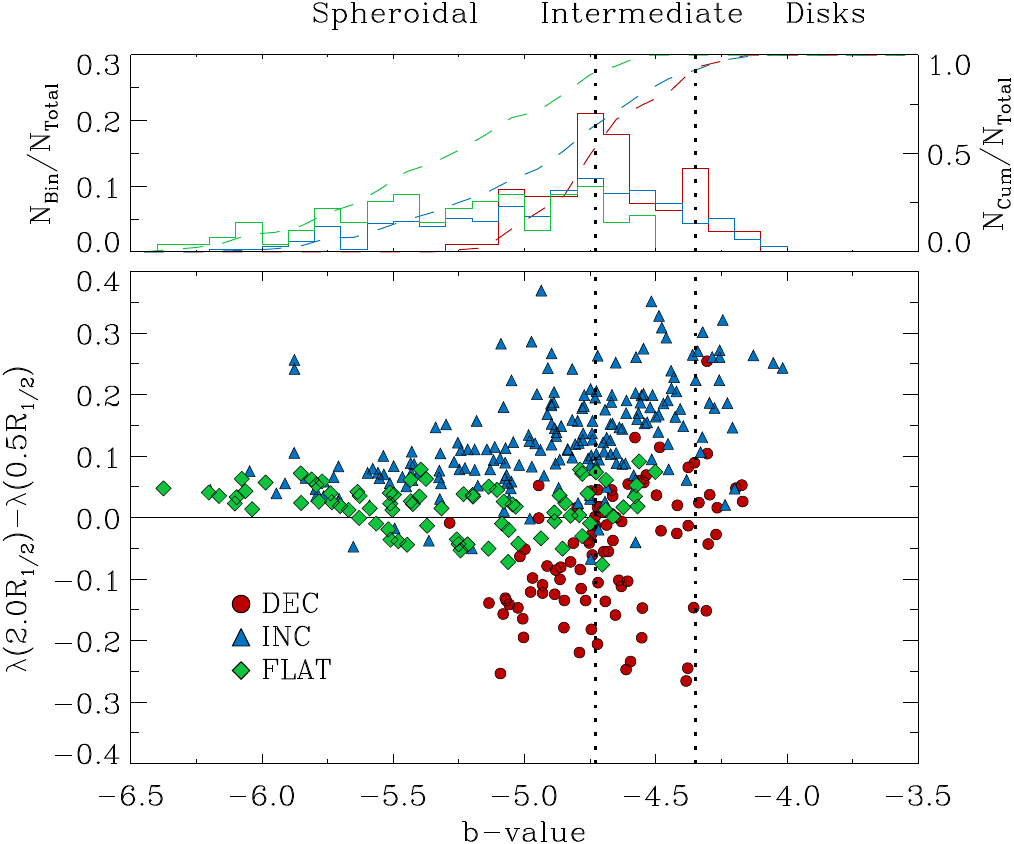}
        \end{center}
        \caption{Gradient in $\lambda(R)$ profile between $2.0R_{\mathrm{1/2}}$ and $0.5R_{\mathrm{1/2}}$ versus the b-value. The sample is subdivided into decreasing profiles (filled red circles), increasing profiles (blue filled triangles), and flat profiles (green squares). The top panel shows the normalised distribution for the three profiles types (solid lines) together with the according cumulative distributions (dashed lines).}
        {\label{fig:fig_4}}
\end{figure}

Fig. \ref{fig:fig_4} displays the connection between the b-value and the gradient $\lambda(\mathrm{2.0R_{1/2}})$-$\lambda(\mathrm{0.5R_{1/2}})$. The sample is split up into increasing, decreasing and flat profiles as given in the legend. The upper panel shows the normalised cumulative (dashed) and differential (solid) distribution of b-values for increasing (blue), decreasing (red) and flat (green) profiles.
\par
The increasers cover a large range from disk-like b-values down to spheroidal-like b-values. Their distribution shows no sharp peaks or substructure indicative of sub populations. Furthermore, the increasers feature the most disk-like morphologies (largest b-value). The flat profiles show a similar distribution to the increasers, however shifted to smaller b-values with a significant overlap. No flats are classified as disk galaxies according to their b-values. Furthermore, the lowest b-values are reached by this class. For the decreasers we find the narrowest distribution of the three classes, reaching up to the extremely disk-like b-values. Around $b=-4.7$ (in the "intermediate" morphological region) the distribution features a peak encompassing $\approx 37\%$ of all decreasers. Hence, decreasing profiles are preferentially in the transition range between pure spheroidals and pure disks, not showing tail to low b-values.
\par
Investigating the overall distribution in the lower panel of Fig. \ref{fig:fig_4}, the maximum absolute value of the gradients increase towards higher, and therefore more disk-like, b-values confirming the finding from \citet{2017MNRAS.467.4540B} and \citet{2018MNRAS.480.3105F}. In general the three classes populate distinct regions in this plane. This separation is, as expected from the definition of the classes, mainly driven by the gradient. Overall, the distribution is not symmetric around $0$, but offset of positive gradients.
\par
Interestingly, we find a significant fraction of decreasing profiles with positive gradient, representing $31\%$ of the decreasing sample, which is counterintuitive. This can be explained by the varying position of the central maximum present for the vast majority of the decreasers. For these galaxies the maximum is at larger radii such that the profiles is indeed decreasing at $\lambda(\mathrm{2.0R_{1/2}})$ but didn't fall below the amplitude at $\lambda(\mathrm{0.5R_{1/2}})$. Furthermore, this is the reason for the overlap of flats and decreasers at $b \approx -4.7$. The overlap between the increasers and decreasers highlights that the introduced classification is useful since it recovers decreasers which are consistent with no or even positive gradient. In addition, we even find $8$ increasers with negative gradients. These galaxies show a common characteristic shape: starting from $0.5R_{\mathrm{1/2}}$ the profiles are falling until they reach a minimum in the range $1.0R_{\mathrm{1/2}}$-$2.0R_{\mathrm{1/2}}$ and subsequently increase. These two examples show that a simple gradient with two sampling points doesn't capture the diversity of the profiles properly and the classification introduced in this study is required.


\section{Dissecting the Formation Pathways of Galaxies using Halo Kinematics} \label{sec:formation_pathway}

In the previous section we analysed the stellar large-scale kinematics at $z=0$ in detail, and their connection to global galaxy properties. However, the simulations enable us to investigate the formation pathways of these galaxies over cosmic time and connect the present-day profile shape to merger characteristics of the individual galaxy and therefore identify imprints of the accretion history in the profiles.

\subsection{Connection to Profile Shape} \label{sec:connection_to_profile_shape}

\begin{figure}
        \begin{center}
                \includegraphics[width=0.4\textwidth]{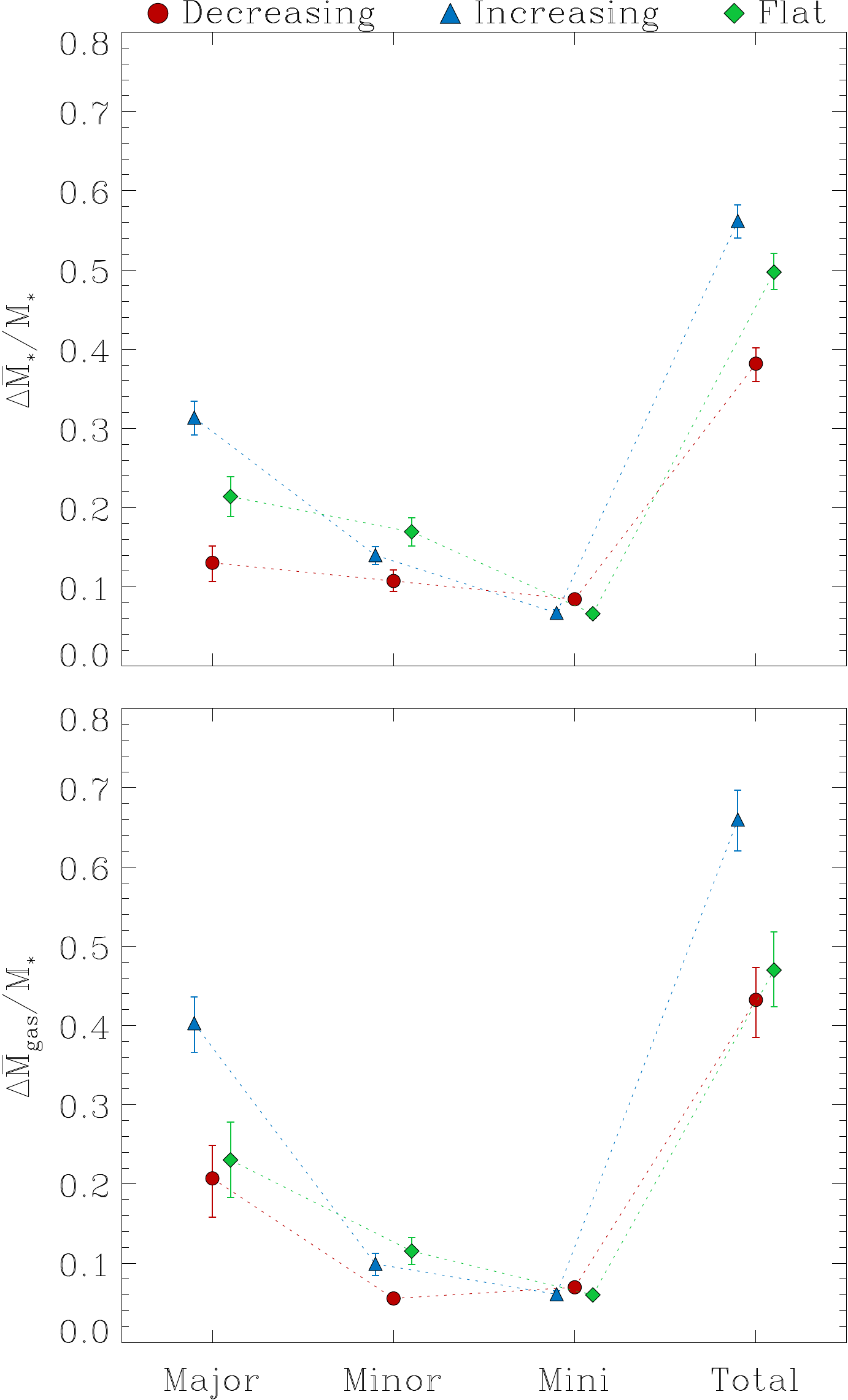}
        \end{center}
        \caption{Both panels shows an averaged quantity related to the accretion history of the galaxies. Different colours and symbols separate the three profile classes decreasing (red filled circles), increasing (blue triangle), and flat (green diamonds). We distinguish between contributions by major, minor, mini and all mergers from left to right. The error associated with each point is derived from $500$ bootstraps and error bars correspond to the $1\sigma$ confidence interval. \textit{Top panel:} Average stellar mass accreted since $z=2.0$ normalised by the stellar mass at $z=0$. \textit{Lower Panel:} Average gaseous mass accreted since $z=2.0$ normalised by the stellar mass at $z=0$.}
        {\label{fig:fig_9}}
\end{figure}

To understand the information about the accretion history encoded in the large scale kinematics, Fig. \ref{fig:fig_9} shows a summary of the merger history for increasing (blue), decreasing (red), and flat (green) profiles condensed into two parameters: The upper panel shows the stellar mass accreted through the various merger types since $z=2$, where $\mathrm{\Delta M_*}$ is normalised by the $z=0$ stellar mass $\mathrm{M_*}$ for each profiles class. Similar to the upper panel of Fig. \ref{fig:fig_9}, the lower panel investigates the amount of gas accreted via the different types of merger since $z=2$. This is a direct measure of the available fuel for potential star formation and the build-up of new dynamically cold components. 
\par
In total, increasers accrete on average $55\%$ of their present day stellar mass through mergers, and therefore $\approx 17\%$ more than decreasers, which accrete $38\%$. Flat profiles gain $50\%$ of their stellar mass via mergers, and hence lie between increaser and decreaser.
\par
The largest difference between the classes is apparent for mass gained through major mergers: While decreasing profiles accrete $13\%$ of their stellar mass through this channel, increaser accrete with $32\%$ almost a factor $3$ more stellar mass via major merger. The flat profile galaxies again reside in the intermediate region, gaining $22\%$ of their stellar mass through major merger. It seems like there is a sequence from increaser over flats to decreaser driven by the total amount of accreted stellar mass and the importance of major merger since $z=2$. 
\par
The amount of stellar mass accreted through minor and mini mergers is similar for the three classes. All classes gain more mass via minor mergers than mini mergers. However, the relative importance of the mini and minor mergers differs strongly between the different classes. Especially, the difference between increasers and decreasers is significant: The mass accreted via mini and major mergers for decreasers is $6\%$ larger than the mass gained through major merger. In contrast, increasers accumulate $11\%$ less via minor and mini merger. Flats are again between the two classes, gaining slightly more stellar mass via mini and minor mergers than through major mergers. Therefore, we conclude, that the merger history of decreasing profiles is dominated by mini and minor mergers that lead to the characteristic profile shape.
\par
Regarding the gaseous mass accreted through mergers for the different classes, we find a similar picture: the decreasing and flat profiles show a very similar behaviour for all merger types, accreting $\approx 45\%$ of their present day stellar mass in gas via merger events and $\approx 7\%$ through each minor and mini merger.Both classes gain $\approx 20\%$ gas mass via major merger. The increasers are clearly separated from the other classes in the total amount of gaseous mass gained through merger events. With $60\%$ gas mass relative to the total stellar mass, this class accretes a factor $1.5$ more gas than decreasers and flats. This behaviour is clearly driven by the major merger component: While the increasers gain a comparable amount of gas through minor and mini merger, they gain a factor $2$ more via major merger than the other classes.
\par
An interesting feature for the increasers is that, when considering the total accreted baryonic mass ($\mathrm{M_*+M_{gas}}$), they assemble more than their present day stellar mass in contrast to the other classes. This suggests, that the increasers exhibit a higher gas-mass fraction and potentially more star formation than the decreasers and flats. Analysing the gas mass fractions confirms this conclusion.
\par
Summarising the conclusion from Fig. \ref{fig:fig_9}, we find distinct differences in the global accretion history of the three profile classes which motivate the following idea for the formation pathway: The merger history of the increasing class is dominated by gas-rich major merging. Depending on the orbital configuration major mergers are believed to enhance the dispersion of the velocity distribution down to the very centre of the halo where the galaxy is located. However, we find a significant amount of gas that is provided by the major merger which potentially can build up a new kinematic cold component as demonstrated in \citet{2017MNRAS.470.3946S}. An increasing $\lambda(R)$ profile is compatible with an exponential disk with a flat rotation curve \citep{2012ApJS..203...17R}. Therefore, we suspect that the gas-rich major mergers for increasing profiles happened rather early before $z \sim 1$, allowing for the build-up of a kinematic cold disk in the centre that is aligned with the stars stripped during the infall. In contrast decreasing profiles are clearly mini and minor merger dominated. Earlier studies have shown that these low-mass satellites, especially the mini merger, never reach the centre of the host halo, getting disrupted in the halo and therefore building up the stellar halo \citep{2017MNRAS.464.2882A,2019MNRAS.487..318K}. Assuming an anisotropic accretion of the satellites, we expect this disruption process to enhance the dispersion in the halo without affecting the central embedded disk. The formation of the decreasing profiles is discussed in more detail Sec. \ref{sec:formation_of_dec}. The flat profiles seem to be in a transition state driven by the ratio of minor+mini to major merger accretion.
\par
\begin{figure}
        \begin{center}
                \includegraphics[width=0.45\textwidth]{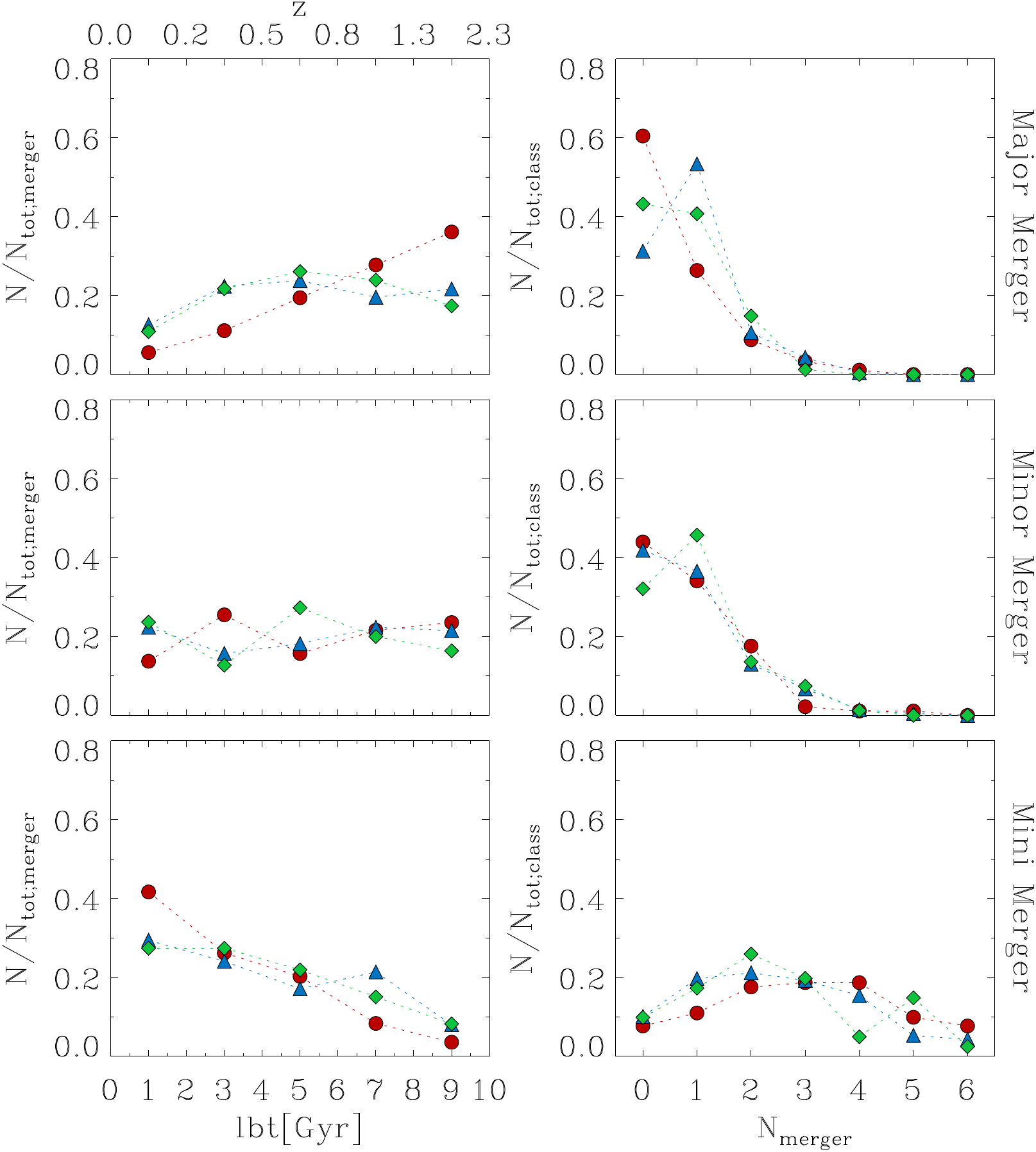}
        \end{center}
        \caption{\textit{Left Column:} The fraction of galaxies that had their last mini (lower panel), minor (central panel) and major (upper panel) merger within a time interval of $\pm 1\mathrm{Gyr}$ around the plotted look-back time (lbt). \textit{Right Column:} The fraction of galaxies that experienced $\mathrm{N_{merger}}$ of a given type. In all panels the different colours and symbols separate decreasing (red circles), increasing (blue triangles) and flat (green diamonds) profiles.
        }
        {\label{fig:plot_further_merger_analyses}}
\end{figure}

In order to elaborate our interpretation of Fig. \ref{fig:fig_9} in more detail, Fig. \ref{fig:plot_further_merger_analyses} shows two more quantities extracted from the merger trees. The right column displays the fraction of galaxies that underwent $\mathrm{N_{merger}}$ major (upper panel), minor (central panel) and mini merger split up into decreasing (red circles), increasing (blue triangles) and flat (green diamonds) profiles.
\par
As expected from Fig. \ref{fig:fig_9}, we find the most sever differences between the classes in the major merger regime. The three classes feature a similar distribution for the fraction to increase towards lower numbers of merger. We find that $61\%$ of the decreasers do not undergo a major merger, and $27\%$ have one merger in their formation history since $z=2$. The increasers show a reversed trend, with the majority of $53\%$ experiencing a single major merger and $30\%$ not having a major merger. The flats are in an intermediate state with approximately the same fraction ($\sim 40\%$) of the sample having $0$ or $1$ major merger. While almost none of the galaxies has more than three minor mergers, $\sim 40\%$ of the sample does not undergo a minor merger at all, similar for the three classes. Furthermore, the distribution for the minor merger exhibits a strong increase towards lower numbers, independent of the profile type. In contrast, in the mini merger case, the distribution declines towards lower and larger numbers with a maximum at $2-3$ mergers similar for all three profile classes.
\par
The left column shows the fraction of galaxies that experienced their last mini (lower panel), minor (central panel) and major (upper panel) merger within a time interval of $\pm 1\mathrm{Gyr}$ around the plotted look-back time (lbt). This means that we only include galaxies that actually experienced a merger of the respective type\footnote{As already discussed in Sec. \ref{sec:Merger_Tree}, in the context of estimating the mass fraction of a merger, determining the precise point in time when a merger happens is notoriously difficult. This is on the one hand due to the algorithm that constructs the merger trees. On the other hand defining a physically meaningful single 'merger time' rather than a period of interaction is also degenerate. Hence, we only focus on overall trends and significant differenced in the left column to not over interpret the artificial definition of a merger time in the merger trees.}.
\par
The major merger (left upper panel) show a contrary trend than the mini mergers (left lower panel): A larger fraction of galaxies experienced an early major merger and no subsequent major merger. This holds true for all three profile classes. While the increasers and flats exhibit a similar distribution the decreaseres again stand out: The outlined trend is significantly stronger for galaxies with decreasing profiles than for galaxies with increasing or flat profiles. About $90\%$ of the decreaseres that experienced a major merger in their formation underwent the merger more than $5 \mathrm{Gyr}$ ago. In contrast, for the increasers and flats this fraction is $\sim 60\%$. Therefore, we conclude that the recent accretion history ($\mathrm{lbt} < 5 Gyr$) of decreaseres that experience a major merger is dominated by mini and minor mergers compared to major mergers. Combined with the finding that only $40\%$ of this class does undergo a major merger at all, this represents a clear and strong signal.
\par
For the minor mergers (central panel) we do not find the same general trend of earlier accretion as for the major merger. The distribution is relatively flat without substantial differences between the profile classes.
\par
The lower left panel shows the general trend that a larger fraction of galaxies had their last mini merger more recently independent of the profile class. The increasers and flats exhibit a very similar evolution with approximately $28\%$ of the sample having their last mini merger in the past $2 \mathrm{Gyr}$, while $8\%$ did not experience a mini merger in the past $9 \mathrm{Gyr}$. The trend is strongest for the decreaseres with $42\%$ experiencing the last mini merger in the past $2 \mathrm{Gyr}$, while only $3\%$ had their last mini merger $9 \mathrm{Gyr}$ ago. Therefore, the fraction of galaxies with $\mathrm{lbt} < 2 \mathrm{Gyr}$ is significantly larger for decreasers than for increasers and flats.
\begin{figure*}
        \begin{center}
                \includegraphics[width=0.9\textwidth]{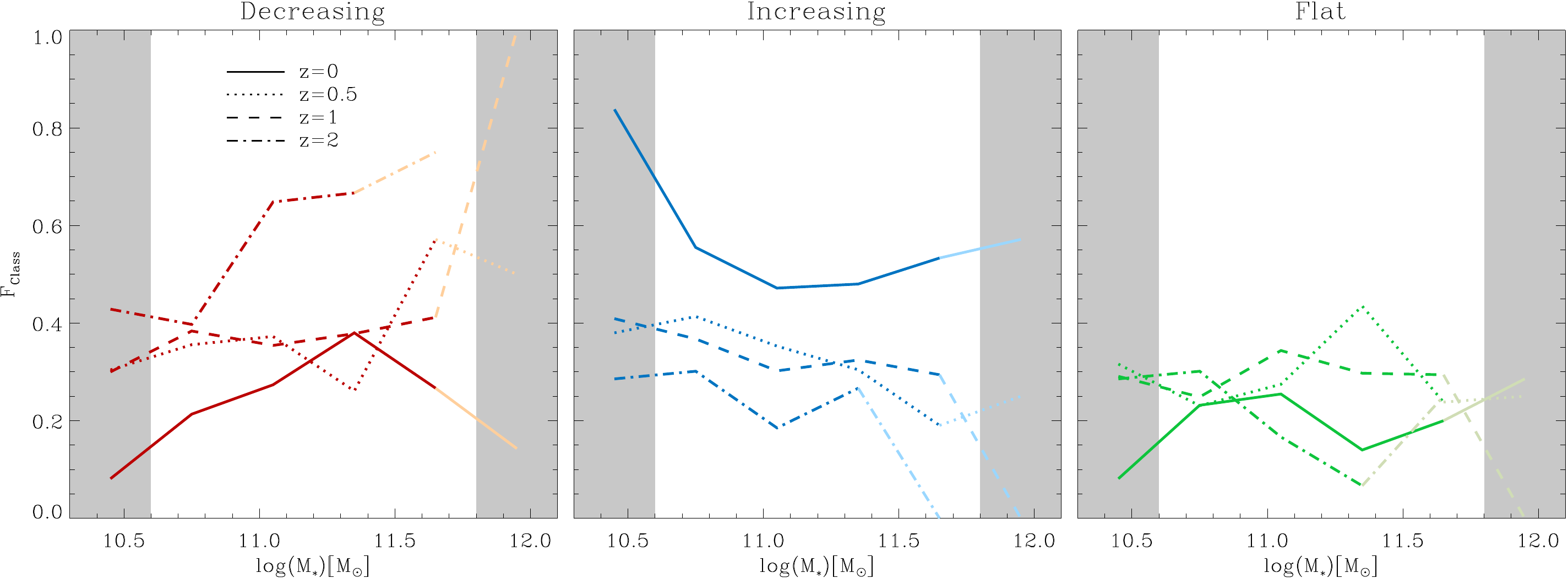}
        \end{center}
        \caption{Each Panel shows the fraction of decreasers (left), increasers (middle), and flats (right) in mass bins with $0.3\mathrm{dex}$ width at $z=0$, $z=0.5$, $z=1.5$, and $z=2.0$. Grey regions mark the bins that are impacted by resolution and the limited box-size. Bins with less than $10$ galaxies are shown in lighter colours.}
        \label{fig:fig_3_redshift}
\end{figure*}

\subsection{Profile Classes at Higher Redshift}

To investigate if the class frequencies evolve over time, we applied our classification scheme at $z=2$, $z=1$, and $z=0.5$ using the same selection criteria as outlined in Sec. \ref{sec:sample_selection}, yielding a samples of $243$, $785$, and $805$ galaxies, respectively. Tab. \ref{tab:redshift_statistics} summarizes the class frequencies at the considered redshifts. We find that the decreasers account for $37\%$ of the $z=2$ sample, while $18\%$ are increasers and $17\%$ are flats. With decreasing redshift, the fraction of decreasers declines to $19\%$ at $z=0$. The largest change in the decreaser fraction happens between $z=2$ and $z=1$. In contrast, the fraction of increasing profiles increases for lower redshift from $18\%$ at $z=2$ to $47\%$ at $z=0$. The fraction of flats does not vary significantly with redshift.
\begin{table}
	\centering
	\caption{The statistical distribution of decreasing, increasing, flat, and unclassified profiles at different redshifts.}
	\label{tab:kin_group}
	\begin{tabular}{cccccc}
		\hline
		 z & Decreasing & Increasing & Flat & Unclass & $\mathrm{N_{gal}}$\\
		\hline
		\hline
		0 & 19\% & 47\% & 18\% & 16\% & 450\\
		0.5 & 27\%& 30\%& 22\% & 21\% & 805\\
		1 & 23\%& 30\%& 22\% & 25\% & 785\\
		2 & 37\%& 18\%& 17\% & 28\% & 243\\
		\hline
	\end{tabular}
	{\label{tab:redshift_statistics}}
\end{table}
\par
At high redshift, a decreasing $\lambda(R)$ profile is the predominant class in contrast to $z=0$, where the increasers comprise half of the total sample. This implies that there has to be an evolutionary path between decreasing and increasing profiles to explain the class frequencies found at $z=0$. Since the decreasing feature is already in place at $z=2$, present-day decreaser must have an accretion history that does not destroy this feature while building up a non-rotating stellar halo, in agreement with our result, that decreasers essentially grew through mini/minor mergers since $z=2$. The origin and nature of decreasing profiles is investigated in more detail in the following chapter. 
\par
In total, $28\%$ of our $z=2$ galaxies are unclassified, representing an increase of $12\%$ in comparison to $z=0$. This is due to a significantly stepper increase of the profiles inside $0.5R_{\mathrm{1/2}}$ for these galaxies and less variation at larger radii, implying that our $z=0$ classification is less distinctive at higher redshift. This reflects the different evolutionary state of the galaxies at $z=2$, strongly dominated by dissipative processes like gas accretion and in-situ star formation.
\par
Fig. \ref{fig:fig_3_redshift} shows the evolution of the distribution in more detail: The class frequency of decreasers (left), increasers (middle), and (flats) is presented in mass bins at the four considered redshifts. It shows that the trends found for the total fractions above is apparent in every mass bin: With decreasing redshift, the frequency of decreasers declines, while the fraction of increasers inclines, independent of mass. Furthermore, we find a general increase in the fraction of decreasers towards higher stellar masses at all redshift, while the increasers show a contrary behaviour. The galaxies with flat profiles do not show an overall trend with redshift.
\par
For lower-masses ($\mathrm{log(M_*/M_{\odot})} < 11$), the most significant drop for decreasers occurs between $z=0.5$ and $z=0$, while for higher masses ($\mathrm{log(M_*/M_{\odot})} > 11$) the decline is strongest already between $z=2$ and $z=1$, as implied by the varying slope with redshift. This behaviour suggests that the process that drives the evolution of $\mathrm{F_{Class}}$ acts at different times for different stellar masses. In contrast, the relatively constant slope for the increasers implies a concurrent rise of the fraction, independent of stellar mass. Only in the lowest mass bin we find a significantly larger jump between $z=0.5$ and $z=0$ than between the other redshifts.

\subsection{Formation of Decreasing Profiles at $z=0$} \label{sec:formation_of_dec}

As shown in Sec.~\ref{sec:connection_to_profile_shape} the formation of present-day decreasing profiles is dominated by low mass-fraction mergers at late times. We suspect that the anisotropic accretion of those satellites and their subsequent disruption in the halo randomises the velocities and hence enhances the dispersion in the halo, while maintaining the central rotating in-situ component. In order to further corroborate this concept, we investigate the evolution of an example galaxy exhibiting a decreasing $\lambda(R)$ profile at $z=0$ in detail.
\begin{figure*}
        \begin{center}
                \includegraphics[width=0.98\textwidth]{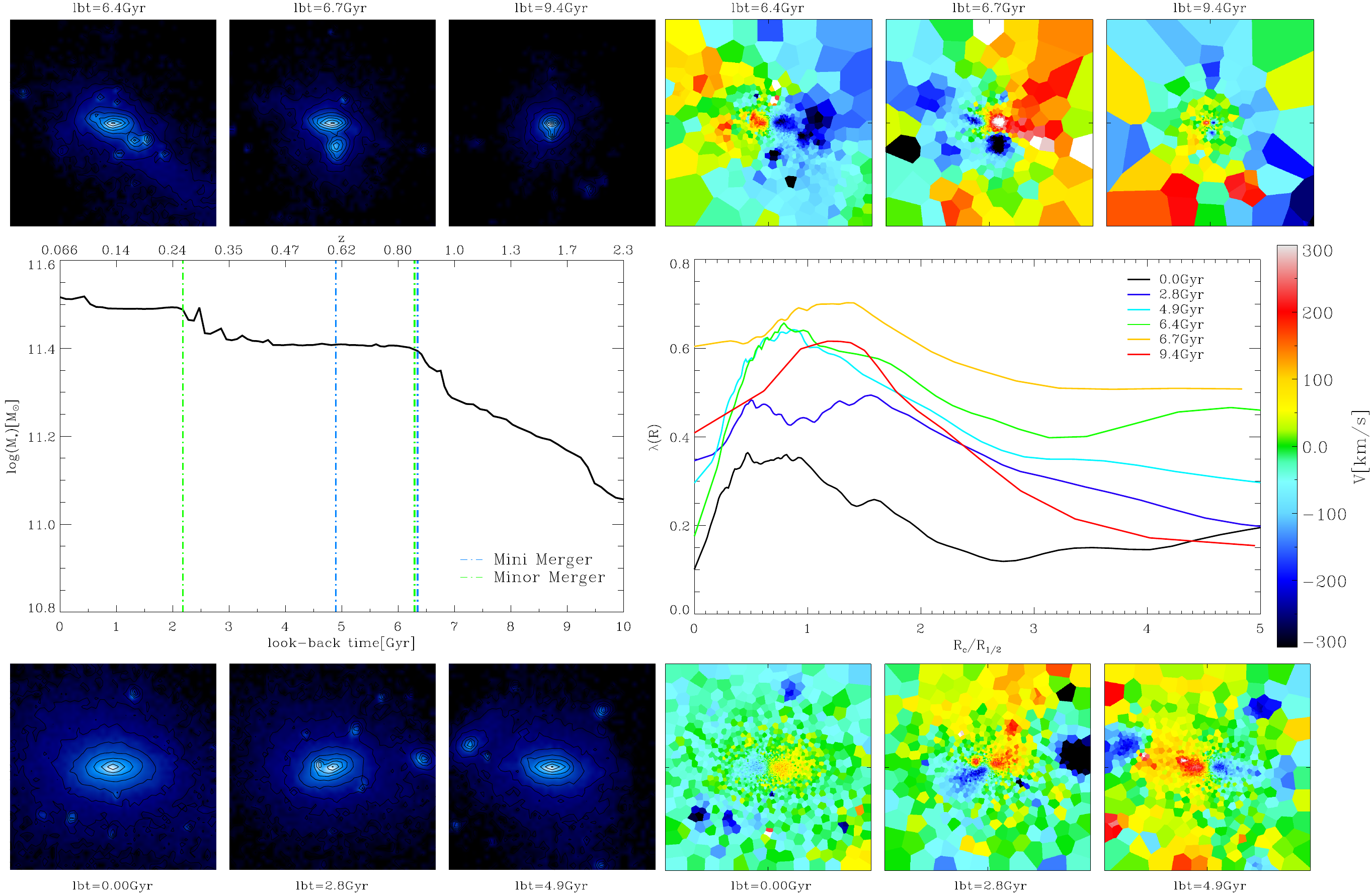}
        \end{center}
        \caption{Visualisation of the temporal morphological and kinematic evolution of the example galaxy with decreasing profile at $z=0$. In the central panel on the left the black solid line illustrates the evolution of the stellar mass with look-back time (lbt). Vertical dashed-dotted lines mark the moment at which a merger is identified. Colours distinguish minor merger (green), and mini merger (blue). The top and bottom panels show the projected stellar density map within a square of $70\mathrm{kpc}$ side length at various distinctive times as given in the panel annotation. The central panel on the right shows the $\lambda(R)$ profile at the same moments in time. Corresponding velocity maps are displayed on the top and the bottom on the right.}
        \label{fig:fig_7_41}
\end{figure*}
\subsubsection{Case Study of a Decreasing Profile}
\begin{figure}
        \begin{center}
                \includegraphics[width=0.43\textwidth]{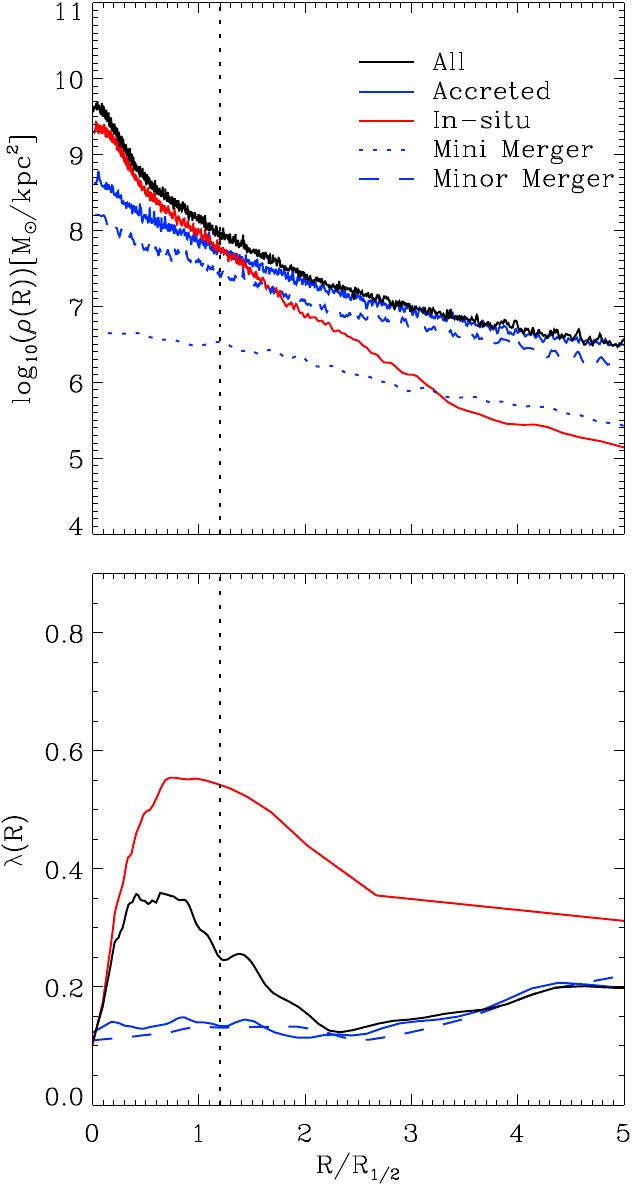}
        \end{center}
        \caption{Further analysis of the example galaxy investigated in Fig. \ref{fig:fig_7_41}. \textit{Upper Panel:} Stellar density profile of the example galaxy at $z=0$ split up into several components. The black solid line shows the profile for all stars while the red and blue solid lines depict the contributions from the in-situ and accreted component constructed from the complete evolution of the galaxy. Blue dashed and dash-dotted curves show the contribution of stars accreted via minor and mini merger since $z=2.0$. \textit{Lower Panel:} $\lambda(R)$ profile split up in the same manner as in the upper panel. Note that it is not possible to construct a $\lambda(R)$ profile for the mini merger stars due to the lack of particles. Vertical lines mark the crossing of the in-situ and accreted density profiles.}
        \label{fig:fig_8_41}
\end{figure}
The left central panel of Fig. \ref{fig:fig_7_41} depicts the evolution of the stellar mass for the example galaxy over cosmic time from $z=2$ down to $z=0$. Vertical dashed lines mark minor (green) and mini (blue) merger as defined in Sec. \ref{sec:Merger_Tree}. This particular example does not experience a major merger in the considered redshift interval. The galaxy experiences two mini and two minor mergers below $z=1$. Relating the example to Fig. \ref{fig:fig_9}, for minor mergers $\mathrm{\Delta M_{*}/M_*=0.29}$, for mini merger $\mathrm{\Delta M_{*}/M_*=0.03}$, and for total merger $\mathrm{\Delta M_{*}/M_*=0.39}$. Therefore, this halo is an example of a minor merger dominated stellar accretion history. For the gaseous component, we find $\mathrm{\Delta M_{gas}/M_*=0.23}$ for minor merger, for mini merger $\mathrm{\Delta M_{gas}/M_*=0.04}$, and for total merger $\mathrm{\Delta M_{gas}/M_*=0.35}$.
\par
The left upper and lower panels show the projected stellar density within a cube of $70 \mathrm{kpc}$, corresponding to $7R_{\mathrm{1/2}}$ at $z=0$, around the galaxy centre at six distinctive times as indicated. The extend of the central disk does not increase significantly for $\mathrm{lbt} < 6.7 \mathrm{Gry}$, while the surrounding stellar halo gets more prominent with time. Although we are not able to trace the orbit of the in-falling satellites in detail, their position suggests that they get accreted from anisotropic directions.
\par
In the velocity maps (right upper and lower panels) we clearly see that the rotating stellar core already present at $\mathrm{lbt} = 9.4 \mathrm{Gyr}$ is visible at all times, and gets not destroyed during the evolution of the galaxy. It gets more pronounced and grows in size towards lower redshift. The accreted satellites are clearly visible in the velocity maps. Especially at $\mathrm{lbt} = 6.4 \mathrm{Gyr}$, we see a coherent velocity structure in the right upper corner which is most probably caused by stripped stars from the in falling satellite. However, the tail is not dense enough to be identified in the density map. The galaxy never shows an ordered rotating motion in the halo, which is reflected in the $\lambda(R)$ profile: At all times the profile is decreasing and maintains its central peak. In general, the entire profile spins down at all radii.
\par
Fig. \ref{fig:fig_8_41} shows various radial properties of the example galaxy at $z=0$, demonstrating the dynamic and kinematic composition of the halo. We select all particles of a merging satellite at the moment when the merger takes place and identify them at $z=0$, for all mergers happening below $z=2$. Furthermore, we follow the complete evolution of the galaxy to identify the stars that have been formed in-situ and stars that have been formed outside and later on got accreted onto the halo. Note that we follow the accreted and in-situ component down to the first identification of the halo, while we only consider merger events below $z=2$. Therefore, the merger components do not fully add up to the total accreted component.
\par
The upper panel of Fig. \ref{fig:fig_8_41} shows the density profile of the different components: total density profile (black solid), accreted (solid blue), in-situ (solid red), mini merger (dotted blue) since $z=2$, and minor merger (dashed blue) since $z=2$ ($n=300$ equal particle bin).
\par
The total profile is well-described by a Sersic profile with a Sersic index of $n=3.0$, consistent with an elliptical galaxy. Decomposing the profile into accreted and in-situ shows that the individual components are closer to an exponential than the total profile. At $\sim 1.2R_{\mathrm{1/2}}$, the accreted and in-situ profiles cross, marking the transition from the in-situ dominated centre to the accretion dominated halo. This is in line with our hypothesis that the decline in the $\lambda(R)$ profile marks a transition of this kind. The stacked mini merger profile contributes only marginally to the total density. In contrast, the minor mergers are a significant contribution to the accreted component. At $\sim 1.7R_{\mathrm{1/2}}$, we find a transition between the minor merger profile and the insitu profile.
\par
In the lower panel of Fig. \ref{fig:fig_8_41} we connect the density profile to the kinematic $\lambda(R)$ profile. The black solid line shows the $\lambda(R)$ profile for all stars, while the other curves decompose the profile in the same manner as in the upper panel. Note that due to the lack of particles it is not possible to construct a $\lambda(R)$ profile for the mini mergers.
\par
The total $\lambda(R)$ profile exhibits a steep incline in the centre within $0.5R_{\mathrm{1/2}}$ up to a peak at $\sim 1.0R_{\mathrm{1/2}}$. Beyond $\sim 1.0R_{\mathrm{1/2}}$ the profile decreases significantly to a minimum at $\sim 2.0R_{\mathrm{1/2}}$, followed by a mild increases beyond that radius. Interestingly, the radial transition range, in which the in-situ and accreted density profiles cross, coincides with the declining section of the total $\lambda(R)$ profile. The $\lambda(R)$ profile for the in-situ component follows a similar shape, however with a much stronger increase in the centre and accordingly a higher peak. At radii larger than $\sim 2.0R_{\mathrm{1/2}}$, the in-situ profile has only very few data points, reflecting the drop in density.
\par
The profile for the accreted component does not show much variation except for a mild increase in the outer radial range. It does not follow the peak of the in-situ profile in the centre, hence mitigating the peak of the total profile to lie below the in-situ profile. For radii larger than $2.5R_{\mathrm{1/2}}$, the $\lambda(R)$ profile of the accreted component coincides with the $\lambda(R)$ profile of the total and the minor merger.
\par
This finding clearly shows that the transition from high rotational support at the peak radius to a less rotational dominated region is driven by the transition from an in-situ dominated inner part to an accretion-dominated halo. In this particular example the halo kinematics are dominated by the stars accreted through minor merger that were stripped from the satellite during the infall.
\par
It also supports the idea, presented by \citet{2018MNRAS.480.3105F}, \citet{2017MNRAS.467.4540B}, and \citet{2014ApJ...791...80A} that the declining profiles indicate an embedded in-situ disk, albeit this disk seems to be rather old. Investigating the age distribution of the in-situ formed stars revealed that $60\%$ of the in-situ stars, which form the highly rotating central disk at $z=0$, were formed before $z=2$, and only $17\%$ below $z=1$. We find a similar behaviour for all our decreasing galaxies at $z=0$, clearly highlighting that the decreasing signal in $\lambda(R)$ really represents a indication for an old embedded stellar disk with a purely mini/minor merger dominated formation history.

\subsubsection{Dynamic versus Kinematic Transition}

In the previous section we showed for an example galaxy with decreasing profile, that the rotating inner part was formed in-situ while the outer halo was formed by only mini and minor mergers. We now test if this is generally the case by investigating where the accreted stars get deposited for a larger sample of galaxies. We quantify this by the half-mass distribution radius $d$: for every single merger event of all decreasing galaxies we select the accreted stars at the moment of the merger and identify them within the halo at $z=0$. The half-mass distribution radius $d$ is then defined to be the radius of a sphere containing half of the total mass of the accreted stars. Therefore, $d$ represents a measure for the radius at which the merger deposited its stars. 
\par
\begin{figure}
        \begin{center}
                \includegraphics[width=0.48\textwidth]{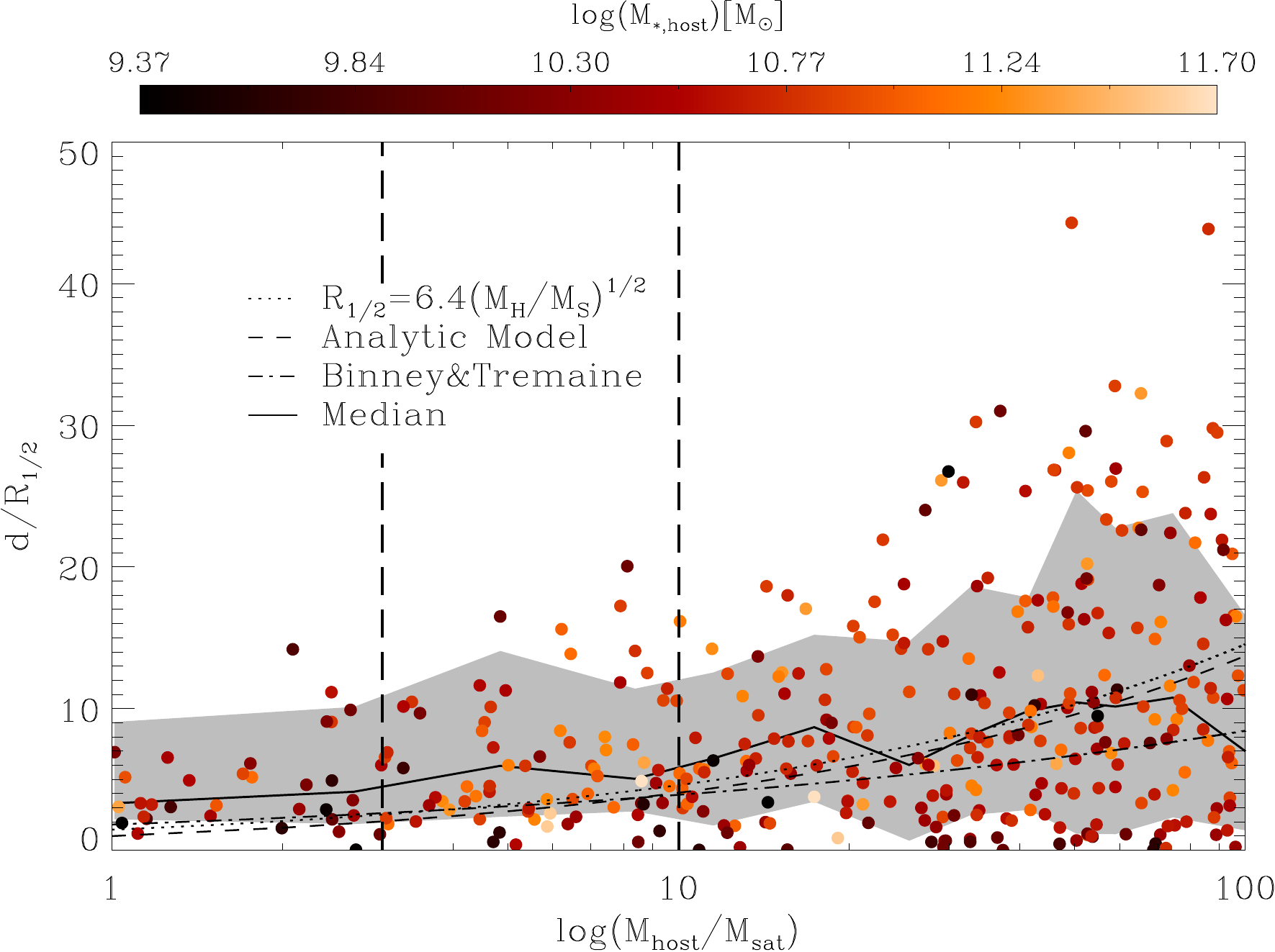}
        \end{center}
        \caption{Half-mass distribution radius $d$ normalized by $\mathrm{R_{1/2}}$ of the host at the moment of the merger event in dependence of the mass-ratio stacked for all merger events experienced by galaxies with decreasing profiles. The colour encodes the stellar mass of the host at the moment of the merger event. Vertical dashed lines visualise the borders between major, minor and mini merger. The black solid line shows the median of the distribution within bins while the grey shaded area mark the corresponding $16\mathrm{th}$ and $84\mathrm{th}$ percentile. The dashed, dotted and dash-dotted curves show predictions from theoretical models extracted from \citet{2019MNRAS.487..318K}.}
        \label{fig:geray_plot}
\end{figure}

Fig. \ref{fig:geray_plot} shows $d$, normalised by $\mathrm{R_{1/2}}$ at the moment of the merger event, versus the mass ratio of the merger, colour-coded according to the stellar mass of the host at the time of the merger. The solid black curve represents the median within bins of $30$ data points, while the grey shaded area marks the $16\mathrm{th}$ and $84\mathrm{th}$ percentile. The three non-solid lines show the prediction from three analytic models presented by \citet{2019MNRAS.487..318K}. Vertical dashed lines visualise the borders between major, minor, and mini merger.
\par
The median curve shows a continuous incline with increasing mass fraction following the behaviour of the analytic models from \citet{2019MNRAS.487..318K} (dashed and dotted). Therefore, lower mass satellites get stripped at larger radii. Similarly, the scatter becomes significantly larger with increasing mass-fraction. While the major merger cover a comparably small range below $\mathrm{d/R_{1/2}}=14$, minor and mini merger populate a significantly broader range. This can be explained by the influence of the orbital configuration of the merger, as found by \citet{2019MNRAS.487..318K}. Investigating a large set of idealised merger simulations with varying orbital configurations and mass-fractions, they showed that major mergers always reach the galaxy at the centre of the halo, regardless of the orbit. In contrast, for minor and especially mini mergers, $d$ is much more sensitive to the orbital configuration and can build up a stellar halo while not affecting the central galaxy. Therefore, minor and mini merger represent a suitable process that leaves the rotating component in the centre intact while building up a kinematic distinct stellar halo. 
\par
To further explore the connection between the kinematic transition and the transition from in-situ to accreted stars we investigate the connection between the position of the maximum of the kinematic profiles $R_{\mathrm{peak}}$ and the density-profile of accreted and in-situ stars for a subset of $43$ decreasers. For this subsample we have access to a well-defined transition radius $\mathrm{R_{tr}}$ that is defined to be the radius at which the density profile of in-situ and accreted stars intersect.
\par
\begin{figure}
        \begin{center}
                \includegraphics[width=0.48\textwidth]{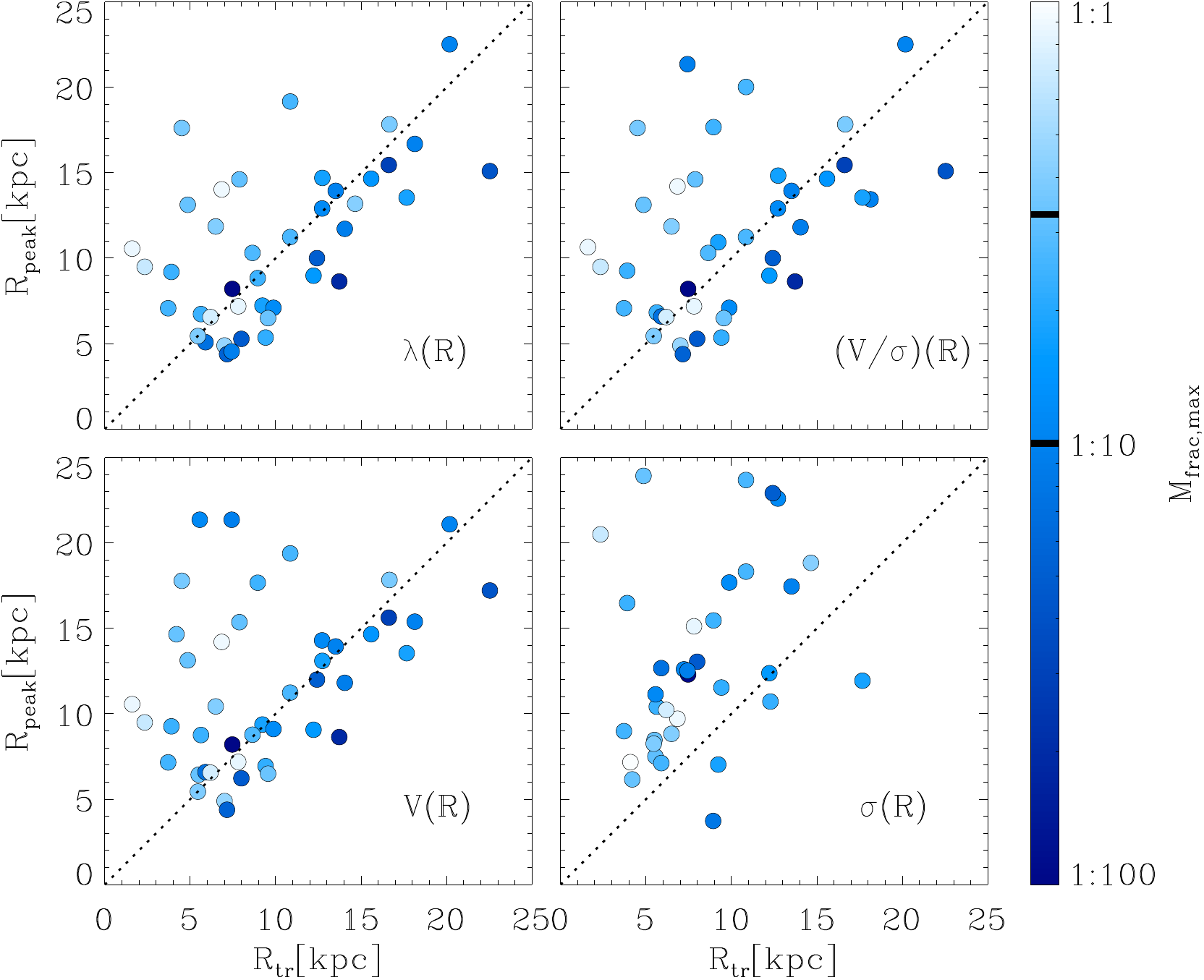}
        \end{center}
        \caption{Radius of the maximum of the $\lambda(R)$ profile $R_{\mathrm{peak}}$ versus the transition radius $R_{\mathrm{tr}}$ at which the accreted component begins to dominate the stellar density profile. The colour encodes the mass-ratio of the most massive merger the galaxy experienced during its formation since $z=2.0$.}
        \label{fig:fig_10.pdf}
\end{figure}

Fig. \ref{fig:fig_10.pdf} correlates $\mathrm{R_{tr}}$ with $R_{\mathrm{peak}}$ for the $\lambda(R)$ profile (upper left), $(V/\sigma)(R)$ profile (right upper), V(R) profile (left lower), and $\sigma(R)$ profile (right lower). The colour encodes the mass-ratio of the most massive merger experienced by the galaxy since $z=2$.
\par
In the upper left panel we can clearly see that there is a apparent trend for $\mathrm{R_{peak}}$ to increase with increasing $R_{\mathrm{tr}}$. The lower right region below the $1$:$1$ line of the diagram is completely unoccupied. Above the $1$:$1$ relation we find a larger orthogonal scatter. The galaxies with the largest distance to the $1$:$1$ line have all experienced major mergers as indicated by the colour. Therefore, the scatter around the $1$:$1$ becomes considerably smaller when only considering the objects that did not undergo a major merger in their evolution. However, a certain scatter in the correlation is expected since in the transition region, where the accreted and in-situ component overlap, the combined kinematics are eminently complex. 
\par
As expected, a very similar behaviour is visible for the $(V/\sigma)$ profile, however with a slightly larger scatter. Decomposing the $(V/\sigma)$ profile in the lower two panels reveals that the velocity is the main driver for the correlation found in the upper panels: While for V the correlation has a comparable scatter around the $1$:$1$ line and with $\mathrm{M_{frac,max}}$, $\sigma$ does not show a correlation close to the $1$:$1$ line at all.
\par
Based on this finding we conclude that, for galaxies with decreasing profiles, the kinematic transition is closely correlated with the transition from an in-situ dominated centre to accretion dominated halo. Furthermore, we predict, that $\mathrm{R_{peak}}$ represents a good estimator for this transition region in many cases. Since $\mathrm{R_{peak}}$ can be observed with a sufficient radial coverage, in contrast to $R_{\mathrm{tr}}$, this represents a meaningful tool to estimate the transition region in observations.

\section{Summary and Conclusion} \label{sec:Summary_and_Conclusion}

We have investigated the kinematic properties of a sample of $450$ galaxies, extracted from the \textit{Magneticum Pathfinder} simulations out to $5 R_{\mathrm{1/2}}$. For this sample we construct $\lambda(R)$ and $(V/\sigma)(R)$ profiles to assess the local kinematic state, and analyse them with regard to galaxy properties and implications on the formation of the galaxies and the halo it resides in.
\par
In a first step we quantify the shape of the $\lambda(R)$ profiles by a simple two-point gradient and compare this to observations from the SAMI, SLUGGS, and ePN.S surveys. Overall, we find an excellent agreement with the SLUGGS and ePN.S observations, while SAMI finds significantly larger gradients also in comparison to the other observational samples. In agreement with the observations, we find negative gradients for a significant fraction of centrally regular rotating objects, with an increaasing fraction when the outer sampling point is shifted to larger radii. This suggests that these galaxies exhibit a rotating component in the centre which is embedded in a less rotationally supported halo, supporting the assumption from \citet{2014ApJ...791...80A} and \citet{2018MNRAS.480.3105F} that these galaxies are disks embedded in a non-rotating stellar halo.
\par
Visually examining the $\lambda(R)$ profiles for our simulated sample reveals three characteristic shapes:
\begin{tasks}
\task Decreasing: The profiles exhibits a central peak in the range $0.5R_{\mathrm{1/2}}$-$2.0R_{\mathrm{1/2}}$.
\task Increasing: The profile is continuously increasing with varying slope until it reaches a plateau.
\task Flat: The profile features only minor variations and stays constant out to the maximally probed radii.
\end{tasks}
Approximately $50\%$ of the sample have an increasing profile, while $20\%$ are decreasing, and $20\%$ are flat. The predominance of of increasing profiles is in agreement with observations by \citet{2017MNRAS.467.4540B}, \citet{2018MNRAS.480.3105F}, and \citet{2014ApJ...791...80A}.
\par
We find that the vast majority of flats are centrally slow rotating galaxies, while the increasing profiles are equally distributed among centrally fast and slow rotators ($50\%$), however showing larger maximum amplitudes in $\lambda(R)$ in the fast rotating regime. Except for one galaxy all the decreasing profiles are classified as centrally fast rotating, reflecting the peak of the profile in the centre.
\par
In order to understand the origin of the different $\lambda(R)$ profile shapes we exploit the full power of the simulation and follow the evolution of our sample through cosmic time. Since galaxy mergers represent one of the major drivers of galaxy evolution at lower redshift, we analyse our sample with respect to merger events. In particular, decreasing profiles show a distinct accretion behaviour: They gain most of their stellar mass via minor and mini merger and not through major merger as the increasing profiles. Analysing the accretion history in more detail for an example galaxy with decreasing profile we find that the central peak in the profile is generated by stars that were formed in-situ at early times, while the slow rotation at larger radii is generated by stars that got accreted in low mass-fraction merger events. This is supported by the finding that at $z=2$ decreasing profiles are, with $37\%$ of the total sample, the predominant class showing that the central peak in the profile is already in place at this time. Furthermore, the radial range of the decline coincides with the transition from in-situ to accreted domination in the density profile. This suggests a formation pathway in which the galaxy merges with low mass galaxies from random direction that get disrupted in the halo due to tidal forces and build up a non-rotating halo while leaving the rotating component in the centre unimpaired.
\par
We support this notion by investigating the distribution of stars being accreted in merger events within the halo at $z=0$. We find that stars that got accreted in minor and mini merger get preferentially deposited at larger radii in comparison to major merger building up the stellar halo. By correlating the position of the peak of the $\lambda(R)$ profile for decreasing profiles with the transition from in-situ to accreted domination in the density profile we predict, that the peak radius represents a good proxy for the transition radius. Since position of the peak of the $\lambda(R)$ profile can be observed, with a sufficient radial coverage, this provides a meaningful tool to estimate the transition radius in observations.
\par
In conclusion we find that galaxies generally show three characteristic local kinematic profiles shapes quantified by $\lambda(R)$. Our conclusions do not change when using $(V/\sigma)(R)$ as a proxy for the local kinematics. We find clear evidence, that the profiles shape encode information about the accretion history of galaxies, especially for the galaxies with decreasing profiles at $z=0$. For these galaxies we can constrain the accretion history to an early in-situ disk formation at $z \geq 2$ and subsequent mini and minor merger that build up the stellar halo. Therefore, decreasing profiles resemble an old disk embedded in a non-rotating accreted halo. Furthermore, $\lambda(R)$ represent a meaningful diagnostic to constrain the radial range where the in-situ or accreted stars dominate the galaxy/halo dynamics. Therefore, our study provides meaningful interpretations and predictions for current and future observations for the formation of galaxies and their stellar halo that can be deduced from the stellar central and halo kinematics.


\section*{Acknowledgments}
We thank Eric Emsellem for helpful discussions and comments. The Magneticum Pathfinder simulations were partially performed at the Leibniz-Rechenzentrum with CPU time assigned to the Project ``pr86re''. This work was supported by the Deutsche Forschungsgemeinschaft (DFG, German Research Foundation) under Germany's Excellence Strategy -- EXC-2094 -- 390783311. FS, RSR, KD, DAF, and SB thank DAAD for financial support.




\bibliographystyle{./mnras}
\bibliography{bibliography} 


\appendix
\section{$\lambda$ versus $V/\sigma$} \label{AppA}

In \citet{2007MNRAS.379..401E} and \citet{2011MNRAS.414..888E} it has been shown that the integrated $\lambda_{R_{1/2}}$ and $(V/\sigma)$ tightly correlate with the form
\begin{ceqn}
\begin{equation}
	\lambda_{\mathrm{R_{1/2}}} = \frac{\kappa (V/\sigma)_{\mathrm{R_{1/2}}}}{\sqrt{1+\kappa^2 (V/\sigma)^{2}_{\mathrm{R_{1/2}}}}}
	{\label{eq:lambda_v_sig}}
\end{equation}
\end{ceqn}
where $\kappa$ is estimated to be $\sim 1.1$. Given that we study local $\lambda(R)$ and $(V/\sigma)(R)$, we want to investigate the connection between $\lambda$ and $(V/\sigma)$ measured locally. 
\par
\begin{figure}
       \begin{center}
                \includegraphics[width=0.48\textwidth]{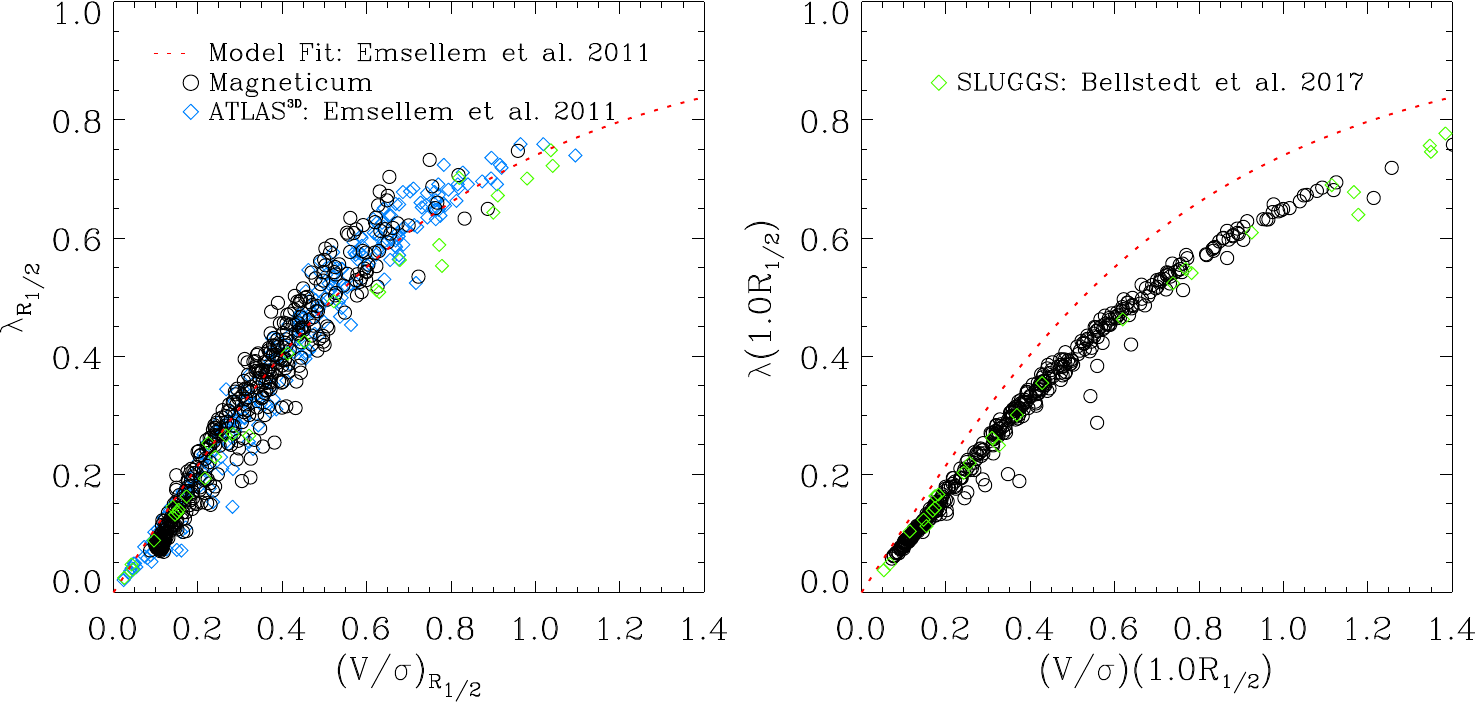}
        \end{center}
        \caption{Correlation between $\lambda$ and $(V/\sigma)$ at a scale of $1.0 \mathrm{R_{1/2}}$. The \textit{left panel} displays the quantities integrated within $\mathrm{R_{1/2}}$. The right panel shows the local quantities calculated within a elliptical shell at $1.0 \mathrm{R_{1/2}}$. Black circles mark the Magneticum galaxies, while the red dashed line is the best fitting relation from \citet{2011MNRAS.414..888E} obtained from dynamical modelling. Blue open diamonds mark observations from the $\mathrm{ATLAS^{3D}}$ survey, while green open diamonds mark observations from the SLUGGS survey extracted from \citet{2017MNRAS.467.4540B}.}
       {\label{fig:appendix}}
\end{figure}
Fig. \ref{fig:appendix} shows the correlation between $\lambda$ and $(V/\sigma)$ integrated within $1 \mathrm{R_{1/2}}$ in the left panel and the local values at $1.0 \mathrm{R_{1/2}}$ on the right. Black circles represent the Magneticum galaxies, while the red dashed lines and the blue diamonds in the left panel mark the best model fit and observations extracted from \citet{2011MNRAS.414..888E}. Observations from the SLUGGS survey extracted from \citet{2017MNRAS.467.4540B} are marked by green diamonds. For the integrated quantities, the simulated galaxies are in good agreement with the theoretical model prediction, however with a slightly steeper slope. A steeper slope, with respect to the model, is also apparent in the observed $\mathrm{ATLAS^{3D}}$ sample. The SLUGGS galaxies exhibit a slightly shallower slope in the high spin regime than than $\mathrm{ATLAS^{3D}}$ and the Magneticum galaxies.
\par
For the locally measured quantities, the correlation is tighter than in the integrated case. This is due to the larger area considered by the integration and therefore more variations in the kinematic maps. In contrast, the calculation of the local values is only based on the area close to an isophote, excluding regions with large variations in the kinematic maps. Although it follows a similar shape, the local correlation exhibit a shallower slope with respect to the model prediction and the correlation found for the integrated values. This can be explained by the way we calculate the local values: Using the circularised radius in Eq. \ref{eq:lambda_rad} explicitly drops the radial dependence of $\lambda$ calculated on an isophote, suggesting $\kappa=1.0$ in Eq. \ref{eq:lambda_v_sig} corresponding to a shallower slope of the relation than for $\kappa=1.1$. We find the same behaviour for the SLUGGS sample.
\par
\begin{figure}
       \begin{center}
                \includegraphics[width=0.45\textwidth]{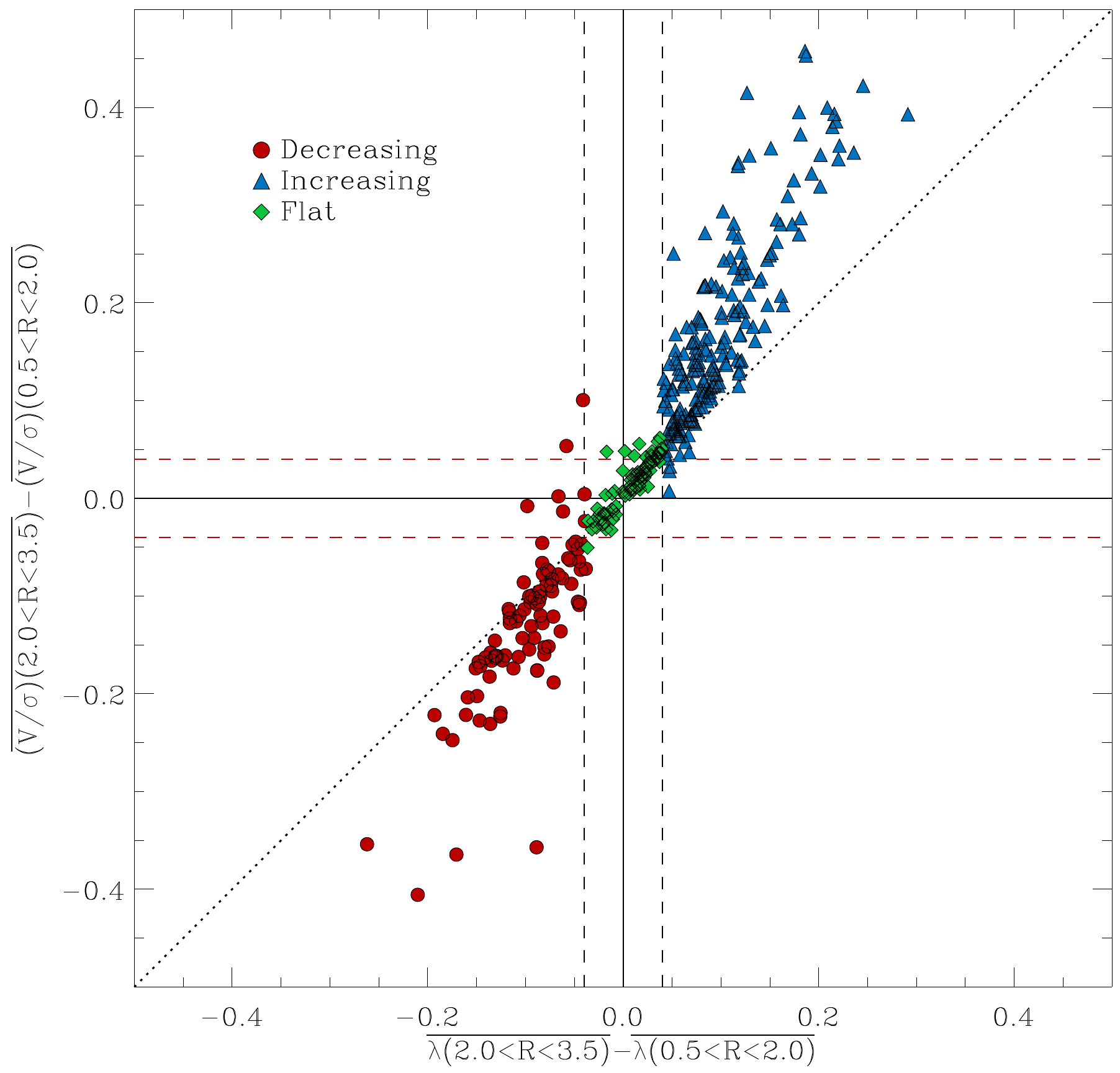}
        \end{center}
        \caption{Gradient determining the classification in decreasing, increasing and flat given in Eqn. \ref{eqn:gradient_classification}. The x-axis displays the gradient for $\lambda(R)$ profile, while the y-axis displays the corresponding gradient of the $(V/\sigma)(R)$ profile. The sample is split up into decreasing (red circles), increasing (blue triangles), and flat (green diamonds) based on the $\lambda(R)$ profile. The vertical dashed lines illustrate the chosen boundaries of the three groups, while the dotted line marks the $1$:$1$ relation.}
       {\label{fig:plot_l_r_v_sig_classification}}
\end{figure}

The profile classification applied in this study is mainly based on the difference between the inner ($0.5 \mathrm{R_{1/2}} <R< 2.0 \mathrm{R_{1/2}}$) and outer ($2.0 \mathrm{R_{1/2}} <R< 3.5 \mathrm{R_{1/2}}$) mean of the $\mathrm{\lambda(R)}$ profile by using a fine tuned threshold of $\pm 0.04$ to separate increasing and decreasing profiles. Fig. \ref{fig:plot_l_r_v_sig_classification} shows this gradient for the $\mathrm{\lambda(R)}$ profile versus the gradient of the $\mathrm{(V/\sigma)(R)}$ profile at the same radii. The $1$-$1$ relation is given by the dotted line, while the vertical dashed lines mark the $0.04$ threshold. Furthermore, the sample is split up into the three profile classes marked by the symbols as indicated in the legend.
\par
As expected from Fig. \ref{fig:appendix}, the two gradients correlate strongly. In the small gradient regime, which is populated by flat profiles, the distribution follows the $1$-$1$ relation closely. Only for more negative and positive gradients, the gradients in the $\mathrm{(V/\sigma)(R)}$ profile diverge to lower and larger values, respectively. This is most probably due the flattening of the relation at larger values seen in Fig. \ref{fig:appendix}, since a stronger gradient suggests larger amplitude values in the flatter regime of the relation. However, it is evident that the chosen threshold value for the $\mathrm{\lambda(R)}$ gradient also represents a meaningful differentiation in the $\mathrm{(V/\sigma)(R)}$ gradients as illustrated by the red dashed line with only few misclassified galaxies. Therefore, we conclude that utilising the $\mathrm{(V/\sigma)(R)}$ profile for classification would not alter the results of the study.

\section{Merger Tree Construction} \label{App:merger_tree}

To construct merger trees we apply the L-HALOTREE algorithm in the postprocessing, which is outlined in the supplementary information of \citet{2005Natur.435..629S}. We give a brief overview of the basic concept here: Initially, halo and subhalo structures are identified for all output snapshots by applying a standard Friends-of-Friends algorithm \citep[FOF, ][]{1985ApJ...292..371D}, assuming a linking length of $0.16$ in combination with an adapted version of the subhalo finder SUBFIND \citep{2009MNRAS.399..497D,2001MNRAS.328..726S}. Due to hierarchical merging a halo can have several progenitors, while it in general only has one specific descendant. Therefore, the algorithm determines the descendant of a halo, which implicitly also yields the progenitor information. For a given halo, the appropriate descendant is identified by tracing the unique particle IDs to the subsequent snapshot and finding all halos that contain particles from this halo. The particles are then counted giving higher weights to the particles that are gravitationally more bound, i.e. have a higher binding energy to the halo under consideration. In this manner, the fate of the inner part of the structure is tracked, which is most resistant with regard to stripping processes that occur during the in-fall into a larger structure. After the weighted counting, the halo with the highest count is selected as descendant. A minor side note: In rare cases it is possible that small structures fluctuate below the detection limit in a subsequent snapshot, however appear again in the following snapshot. To deal with this cases the algorithm allows halos to skip snapshots instead of losing track of the structure.


\bsp	
\label{lastpage}
\end{document}